\documentclass[useAMS,usenatbib]{mn2e_personal}

\usepackage{graphicx}

\title[GASPHER: a key for Lagrangian gas-dynamics]{An approach for solving the boundary free edge difficulties in SPH modelling: application to a viscous accretion disc in close binaries}
\author[G. Lanzafame]{G. Lanzafame\thanks{E-mail:
glanzafame@oact.inaf.it}\\
INAF - Osservatorio Astrofisico di Catania, Via S. Sofia
              78 - 95123 Catania, Italy\\}
\begin{document}

\date{Accepted -------. Received -------; in original form -------}

\pagerange{\pageref{firstpage}--\pageref{lastpage}} \pubyear{2009}

\maketitle

\label{firstpage}

\begin{abstract}
  Adaptive spatial domains are currently used in Smooth Particle Hydrodynamics (SPH) with the aim of performing better spatial interpolations, mainly for expanding or shock gas dynamics. In this work, we propose a SPH interpolating Kernel reformulation suitable also to treat free edge boundaries in the computational domain. Application to both inviscid and viscous stationary low compressibility accretion disc models in Close Binaries (CB) are shown. The investigation carried out in this paper is a consequence of the fact that a low compressibility modelling is crucial to check numerical reliability. \\
  Results show that physical viscosity supports a well-bound accretion disc formation, despite the low gas compressibility, when a Gaussian-derived Kernel (from the Error Function) is assumed, in extended particle range - whose Half Width at Half Maximum (HWHM) is fixed to a constant $h$ value - without any spatial restrictions on its radial interaction (hereinafter GASPHER). At the same time, GASPHER ensures adequate particle interpolations at the boundary free edges. Both SPH and adaptive SPH  (hereinafter ASPH) methods lack accuracy if there are not constraints on the boundary conditions, in particular at the edge of the particle envelope: Free Edge (FE) conditions. In SPH, an inefficient particle interpolation involves a few neighbour particles; instead, in the second case, non-physical effects involve both the boundary layer particles themselves and the radial transport. \\
  Either in a regime where FE conditions involve the computational domain, or in a viscous fluid dynamics, or both, a GASPHER scheme can be rightly adopted in such troublesome physical regimes. \\
  Despite the applied low compressibility condition, viscous GASPHER model shows clear spiral pattern profiles demonstrating the better quality of results compared to SPH viscous ones. Moreover a successful comparison of results concerning GASPHER 1D inviscid shock tube with analytical solution is also reported.
\end{abstract}

\begin{keywords}
accretion, accretion discs -- hydrodynamics -- methods: numerical, N-body simulations -- stars: binaries: close, dwarf novae, cataclysmic variables
\end{keywords}

\section{Introduction}

  In its original version \citep{b35,b36,b42} SPH was formulated adopting a constant particle smoothing length (spatial smoothing resolution length or resolving power) $h$, where the adopted interpolation Kernel works, to perform free  Lagrangian gas dynamics. ASPH methods are currently adopted with the aim of performing better spatial interpolations mainly in expanding or in shock gas dynamics \citep{b8,b13,b3,b36,b45,b46,b12,b15,b52,b48,b39,b54,b31}. High physical viscosity accretion discs are well-bound structures around the primary compact star even in low compressibility conditions \citep{b19,b20,b42,b28}.

  In order to build up a well-bound accretion disc in inviscid conditions, the ejection rate at the disc's outer edge must be at least two or three times smaller than the accretion rate at the disc's inner edge. Whenever this condition is fulfilled, the disc's outer edge, as well as the whole disc, does not disperse in spite of high pressure forces which are also dependent on the gas compressibility: $- \nabla p/\rho = - (\gamma - 1) \nabla (\rho \epsilon)/\rho$. Therefore, low compressibility gases are naturally more easily sensitive to the loss of blobs of gas at the disc's outer edge itself, towards the empty external space, if the gravitational field is not able to keep disc gas in the gravitational potential well. Such effects are enhanced and strongly evident in inviscid conditions \citep{b34,b23} and the moderate contribution of artificial viscosity terms does not prevent such effects. Such a viscosity does not work like a true physical one since it operates only when different fluid components approach each other, being zero during fluid particle repulsion.

  High compressibility gas dynamics does not allow us to distinguish the truth regarding whether a technique is able to perform a correct fluid dynamics. In fact, in such a modelling, accretion discs would be formed anyway even in physically inviscid conditions \citep{b34,b23} and the role of Kernel choice and of its resolving power are hidden. To stress such an idea, in this work a low compressibility $\gamma = 5/3$ polytropic index is adopted throughout, working with the same binary system parameters such as stellar masses and their separation and adopting the largest value ($\alpha_{SS} = 1$)as for the Shakura and Sunyaev viscosity prescription.

  In this paper, physically inviscid and viscid disc models are shown, where a more suitable Gaussian-derived Kernel formulation, as far as both transport mechanisms and expanding or collapsing gas dynamics are concerned, is adopted. Throughout the accretion disc models, the same supersonic mass transfer condition at L1 are adopted.

  The numerical scheme here adopted, as any other numerical method, is characterized by the assumed spatial smoothing resolution length $h$. The mass and angular momentum radial transport is also affected by the SPH particle smoothing resolution length $h$. Too small $h$ values prevent the radial transport, while large $h$ values produce a too effective radial transport of matter towards the centre of the gravitational potential well, as well as of angular momentum toward the disc's outer edge. A large $h$ ensures a high particle overlapping (interpolation) but at the same time it produces a strong particle repulsion rate due to pressure forces especially in low compressibility regimes on the disc's outer FE. On the contrary, a too small smoothing resolution length $h$ compromises any fluid dynamic behaviour and shock handling. The artificial viscosity term prevents spurious heating and handles shocks as a "shock capturing method". The artificial viscosity is a function of the smoothing resolution length itself or of some kind of spatial length. A too small $h$ value does not prevent particle interpenetration, destroying any fluid behaviour, because of lack of artificial viscosity. \citet{b34,b23,b18,b56} and \citet{b28} discuss what we statistically define as a well-defined and bound accretion disc. As far as the numerical resolution is concerned, a number of disordered neighbour particles of the order of 10 (more or less) is considered, in principle, the minimum number of neighbours in order to achieve an adequate 3D numerical interpolation, although a number of neighbours larger than 30 is currently adopted to achieve a higher accuracy. This is the criterion we adopted to define a well-bound accretion disc. Lesser neighbours for each particle are considered an unsuitable number as far as both interpolation efficiency and disc binding into the primary's gravitational potential well are concerned.

  In the next sections, after discerning the artificial and the turbulent physical viscosities, we describe how ASPH techniques work and their limits when the viscous transport and/or FE conditions are involved, as well as why GASPHER could be a solution. In particular, in \S2 we compare how artificial and turbulent physical viscosities differently work; in \S3 we show how GASPHER works and why it does not suffer of some SPH and/or ASPH lack. At last, in \S5 we report 3D accretion disc results showing some interesting features in our viscous simulations, whilst in \S6 we discuss on the accuracy of SPH-derived techniques. In the Appendix, after showing the mathematical background underlying SPH-derived schemes (for readers knowing how SPH and ASPH work, this mathematical section can be easily skipped without any difficulty, being instead essential for others), we also compare results of GASPHER, SPH and ASPH non viscous 1D and 2D selected tests. A comparison with analytical solutions is also given, whenever it is possible.

%
%__________________________________________________________________
%
%

\section{The artificial and the turbulent physical viscosities}

  In our physically viscous disc modelling, the Shakura and Sunyaev prescription  \citep{b50,b70,b51} is adopted with the largest $\alpha_{SS} = 1$ value to stress numerical reliability of results \citep{b28,b56}. The SPH formulation of viscous contributions in the Navier-Stokes and energy equations has been developed by \citet{b10,b11}. These goals are not obtained by artificial viscosity which is, however, introduced in both models to resolve shocks numerically and to avoid spurious heating. Artificial viscosity vanishes when the limit value of the particle interpolation domain goes to zero. \citet{b32,b7,b44} and \citet{b47} demonstrated that the linear component of the artificial viscosity itself, in the continuum limit, yields a viscous shear force. In particular, the last two authors have explicitly formulated such an artificial viscosity contribution in the momentum and energy equations. Moreover, \citet{b44} and  \citet{b47} found an analogy between the shear viscosity generated by the linear artificial viscosity term and the well-known Shakura and Sunyaev shear viscosity, in the continuum limit. SPH method, like other finite difference schemes, is far from the continuum limit; moreover we need the quadratic ($\beta_{SPH}$, Von Neumann-Richtmyer-like viscosity) artificial viscosity term to handle strong shocks. Linear $\alpha_{SPH}$ and quadratic $\beta_{SPH}$ artificial viscosity terms (usually $\sim 1$ and sometimes, in some specific cases, $< 1$) are chosen $= 1$ and $= 2$, respectively. In the viscous models, the viscous force contribution is represented by the divergence of the symmetric viscous stress tensor in the Navier-Stokes equation. A symmetric combination of the symmetric shear tensor times the particle velocity has been added to the energy equation as a viscous heating contribution. The bulk physical viscosity contribution has not been considered for the sake of simplicity.

  Artificial and turbulent physical viscosities are independent from each other. The artificial viscosity terms should be smaller than the physically viscous ones, otherwise the physical viscosity role would be negligible. The relevance of viscous forces could be even comparable to the gas pressure forces, especially if $\alpha_{SS} = 1$ \citep{b28,b56}. An analytical formulation, describing the numerical artificial viscosity coefficient, is reported in \citet{b34}: $\nu_{SPH} = c_{s} h$, where $c_{s}$ is the sound velocity. According to such a definition, its ratio with the Shakura-Sunyaev viscosity coefficient $\nu_{SS} = \alpha_{SS} c_{s} H$ is: $\nu_{SPH}/\nu_{SS} = h/(\alpha_{SS} H)$, for each SPH particle. For $h/H = 5 \cdot 10^{-2}$, where $H$ is the scale-height of the disc, $\nu_{SPH}/\nu_{SS} \sim 5 \cdot 10^{-2}/\alpha_{SS}$. This implies that the role of artificial viscosity could be significant, compared to the physical viscosity role, if small $\alpha_{SS}$ and large $h$ values are taken into account. According to  \citet{b44} and to  \citet{b47} the shear viscosity $\nu_{SPH} \sim 0.1 \alpha_{SPH} c_{s} h$ with $\nu_{SS} = \alpha_{SS-artif} c_{s} H$. According to their results, the numerical artificial viscosity coefficient $\nu_{SPH}$ is even smaller if $\alpha_{SPH} \sim 1$. In fact, the ratio $\nu_{SPH}/\nu_{SS} \simeq 0.1 h/(\alpha_{SS} H)$. Hence, for $H/h \sim 10 \div 20$, $\nu_{SPH}/\nu_{SS} \simeq (5 \cdot 10^{-3} \div 10^{-2}) /\alpha_{SS}$. This implies that, the role of artificial viscosity can be comparable to the role of a very low physical viscosity, because of the correlation between the SPH artificial viscosity parameter $\alpha_{SPH}$ and the Shakura-Sunyaev viscosity parameter $\alpha_{SS-artif}$ is: $\alpha_{SS-artif} \sim 0.1 \alpha_{SPH} h/H$ without any bulk viscosity contribution and supposing gas incompressibility ($\nabla \cdot \bmath{v} = 0$). Notice that, according to these correlations, the Shakura-Sunyaev parameter $\alpha_{SS-artif}$ (non zero only for approaching particles) is not the Shakura-Sunyaev viscous parameter $\alpha_{SS}$ for physically viscid gases, but the transformation of the artificial viscosity term into the Shakura-Sunyaev formalism. Such results show that the gas compressibility has a relevant role since the physical viscosity mainly works when the density varies on a length-scale of the order of the velocity length-scale, not only as a bulk viscosity, but also as a shear viscosity. Moreover, notice that the assumption of an adaptive $h$ SPH or a constant $h$ SPH could also have a role both in artificial viscosity and in physical viscosity roles. These results show that the role of a fully viscous fluid dynamics is still far from any conclusion and that physical assumptions as well as numerical hypotheses and boundary conditions are also determinant.

%
%__________________________________________________________________
%
%

\section{Viscous fluid dynamics equations}

  As for viscous gas hydrodynamics, the relevant equations to our model are:

\begin{equation}
\frac{d\rho}{dt} + \rho \nabla \cdot \bmath{v} = 0 \hfill \mbox{continuity equation}
\end{equation}

\begin{eqnarray}
\frac{d \bmath{v}}{dt} & = & - \frac{\nabla p}{\rho} + [-2 
\bmath{\omega} \times \bmath{v} + \bmath{\omega} \times 
(\bmath{\omega} \times \bmath{r}) - \nabla \Phi_{grav}] + \nonumber \\
& & \frac{1}{\rho} \nabla \cdot \bmath{\tau} \ \ \ \ \ \hfill \mbox{Navier-Stokes momentum equation}
%\nonumber
\end{eqnarray}

\begin{eqnarray}
\frac{d}{dt} \left( \epsilon + \frac{1}{2} v^{2}\right) = - \frac{1}{\rho} \nabla \cdot \left( p \bmath{v} - \bmath{v} \cdot \bmath{\tau} \right) + \bmath{g} \cdot \bmath{v} \nonumber
\end{eqnarray}

\begin{equation}
\hfill \mbox{energy equation}
\end{equation}

\begin{equation}
p = (\gamma - 1) \rho \epsilon \hfill \mbox{perfect gas equation}
\end{equation}

\begin{equation}
\frac{d \bmath{r}}{dt} = \bmath{v} \hfill \mbox{kinematic equation}
\end{equation}

  The most of the adopted symbols have the usual meaning: $d/dt$ stands for the Lagrangian derivative, $\rho$ is the gas density, $\epsilon$ is the thermal energy per unit mass, $\Phi_{grav}$ is the effective gravitational potential generated by the two stars and $\bmath{\omega}$ is the angular velocity of the rotating reference frame, corresponding to the rotational period of the binary system. Self-gravitation has not been included, as it appears irrelevant. The adiabatic index $\gamma$ has the meaning of a numerical parameter whose value lies in the range between $1$ and $5/3$, in principle. $\bmath{\tau}$ is the viscous stress tensor, whose presence modifies the Euler equations for a non viscous fluid dynamics in the viscous Navier-Stokes equations.

\section{Classical SPH Kernel and particle smoothing resolution length}

  In its original formulation \citep{b35,b36,b42} Gaussian Kernels $W_{G,ij} = W(r_{ij},h)$ as:

\begin{eqnarray}
W_{G,ij} & = & \frac{1}{h \sqrt{\pi}} e^{- r_{ij}^{2}/h^{2}} \ \ \ \ \ \ \ \ \ \ \ \ \ \ \ \ \ \ \ \ \ \ \ \ \ \ \ \ \ \ \hfill \mbox{\normalsize , in 1D} \nonumber \\
 & & \\
W_{G,ij} & = & \frac{1}{h^{3} \pi^{3/2}} e^{- r_{ij}^{2}/h^{2}} \ \ \ \ \ \ \ \ \ \ \ \ \ \ \ \ \ \ \ \ \ \ \ \ \ \ \ \hfill  \mbox{\normalsize , in 3D} \nonumber 
\end{eqnarray}

have been adopted in SPH, where $r_{ij} = |\bmath{r}_{ij}| = |\bmath{r}_{i} - \bmath{r}_{j}|$ represents the module of the radial distance between particles $i$ and $j$. Also, an example of "Super Gaussian Kernel" \citep{b36} has also been described. Even a factorization of Gaussian Kernels for each dimension has also been adopted \citep{b52,b48} in an ASPH formulation, adopting 3D ellipsoid Kernel geometry to achieve a higher accuracy, according to an anisotropic $\nabla p$-dependent spatial particle concentrations; or according to the mean particle spacing, as it varies in time, space, and direction around each particle \citep{b31}. Kernels based on cubic splines since the end of the 80's \citep{b42,b36} have also widely been adopted. Typically, in 3D, such cubic spline Kernels $W(r_{ij},h)$ are in the form:

\begin{equation}
W_{3S,ij} = \frac{1}{\pi h^{3}} \left\{ \begin{array}{ll}
1 - \frac{3}{2} q_{ij}^{2} + \frac{3}{4} q_{ij}^{3} & \textrm{if $0 \leq q_{ij} \leq 1$}\\
\frac{1}{4} (2 - q_{ij})^{3} & \textrm{if $1 \leq q_{ij} \leq 2$}\\
0 & \textrm{otherwise,}
\end{array} \right.
\end{equation}

where $q_{ij} = r_{ij}/h$.

\section{GASPHER an alternative way for Kernel and smoothing length}

\subsection{Lack of SPH and ASPH in FE conditions}

  The hidden problem is whether ASPH, as previously formulated, are effective whatever is the compressibility regime considered, especially when FE conditions are adopted on the edges of the particle envelope. High compressibility gas dynamics prevents us from distinguishing the truth regarding whether a SPH-like technique is able to perform a correct fluid dynamics, since accretion discs would be formed anyway even in physically inviscid conditions. In this case, the roles of the Kernel choice and of its resolving power are hidden. Gas loss effects in low compressibility conditions naturally develop, especially at the disc's outer edge, because of the pushing action towards the outer space of particles just below the disc's surfaces and below the disc's outer edge, if the gravitational field is not able to keep gas particles in the gravitational potential well. In ASPH interpolation particle domains swell at both free edges (inner and outer). Normally, in an accretion disc, the density is a decreasing function of the radial distance from the central star. This implies that particle adaptive $h$ should decrease towards the inner disc bulk, without any restriction imposed on the number of particle neighbours. Problems deriving from the inadequacy of artificial viscosity role and the particle interpolation/interpenetration could be relevant. Even the choice of a threshold value for $h_{min}$ as a lower limit would be arbitrary and no differences would appear in results compared to classical SPH results, adopting the same $h = h_{min}$. If ASPH is adopted, even restricting the particle neighbours to a fixed number in its conservative form, the behaviour of $h$ for each particle is contrary, swelling also within the disc bulk and producing enhanced gas loss effects at the disc's outer edge and on the disc surfaces, in spite of the viscosity eventually introduced, as well as a draining effect of the disc's inner edge toward the central compact star due to a stressed radial transport.

  Whenever and wherever spatial isotropy and homogeneity hold, a modulation of spatial smoothing resolution length does not affect results, in principle, in so far as $h$ is large enough to prevent particle interpenetration and neighbour particles are enough to allow good interpolations. However, the situation is rather different if spatial gradients exist. 

  It is quite normal that a smaller threshold limit $h_{min}$ is imposed on particle $h$ because problems on the ineffectiveness of artificial viscosity in handling shock fronts would arise if $h \rightarrow 0$, together with a too short time step computed according to the Friedrich-Courant-Lewy conditions. Artificial viscosity vanishes when the limit value of the particle interpolation domain goes to zero, due to the fact that in its analytical expression it is linearly dependent on the smoothing length $h$. Its role, limited to a filing effect, should not be dominant compared to gas pressure terms. Therefore such condition is fully altered if, according to eqs. (18, 19), $\eta_{ij} \simeq 1$, see App. A. Other formulations of artificial viscosity, depending on particle mutual distance $r_{ij}$ \citep{b38}, do not modify the problem. Therefore, ASPH results would be deeply influenced by a dominant role of artificial terms if such a condition is mostly realized in sonic and subsonic regimes and/or in progressive turbulent rarefying regimes. Some authors \citep{b43,b48} handle artificial viscosity switching it off, especially in low density conditions, when particle $h$ increases or in high temperature conditions when particle sound velocity is subsonic. As a result, the switching on/off of the artificial viscosity limits its role, but low density sonic and subsonic conditions stay still be critical.

  In both situations, the problem of a correct hydrodynamics involves not only the bulk of the gas structure in the computational domain, but mainly the physics of the FE of the computational domain. In particular the outer one for gas expansion problems and the inner one for collapse problems.

  As for physically viscous ASPH simulations, mass and angular momentum transport are deeply affected by the $h$ particle smoothing resolution length. We expect a higher particle transport when particle $h$ statistically increases and the opposite effect when $h$ statistically decreases. In a low compressibility regime, \citet{b28,b56} showed that physical turbulent viscosity hampers particle repulsion, due to pressure forces, contributing to accretion disc consistency and limiting particle loss at the disc's outer edge. However, if particle smoothing resolution length $h$ increases in ASPH, and radial transport becomes unnaturally too much effective, the opposite effect arises so much that the inner edge of the disc could be indefinite.

\subsection{The Kernel of GASPHER: comparison to other Kernels}

  In GASPHER modelling, a radial Gaussian-derived Kernel, related to the well-known "Error Function" with a constant smoothing length $h$ equivalent to its HWHM is considered:

\begin{equation}
W_{ErF,ij} = \left\{ \begin{array}{ll}
\frac{2}{\pi^{1/2} h} e^{- r_{ij}^{2}/h^{2}} & \textrm{, in 1D}\\
\frac{1}{\pi^{3/2} h r_{ij}} e^{- r_{ij}^{2}/h^{2}} & \textrm{, in 2D}\\
\frac{1}{2 \pi^{3/2} h r_{ij}^{2}} e^{- r_{ij}^{2}/h^{2}} & \textrm{, in 3D.}
\end{array} \right.
\end{equation}

In such a Kernel we stress that its interpolation radial extension is unlimited, although its typical smoothing length $h$ is spatially and permanently constant. In GASPHER, to collect an adequate particle neighbours number is not a problem because of the unlimited spatial extension of its Kernel. In the continuum limit, the three interpolation Kernels give the same interpolation integrals for 1D flows, as well as the last two Kernels give the same interpolation integrals for 2D flows.

  The origin of this Kernel function relies in the well known "Error Function":

\begin{equation}
ErF(x) = \frac{2}{\sqrt{\pi}} \int_{0}^{x} e^{- t^{2}} dt,
\end{equation}

whose "Complementary Error Function" is:

\begin{equation}
ErFC(x) = 1 - ErF(x) = \frac{2}{\sqrt{\pi}} \int_{x}^{\infty} e^{- t^{2}} dt.
\end{equation}

  For $x = 0$,

\begin{equation}
ErFC(0) = 1 - ErF(0) = \frac{2}{\sqrt{\pi}} \int_{0}^{\infty} e^{- t^{2}} dt = 1.
\end{equation}

  For $x = 0$, $ErFC(0)$ equals the zero order Gaussian integral:

\begin{equation}
I_{0} = \int_{0}^{\infty} e^{- \xi t^{2}} dt = \frac{1}{2} \frac{\sqrt{\pi}}{\xi}.
\end{equation}

  In performing 3D integral,
  
\begin{eqnarray}
\int W_{ErF,ij} d^{3} r_{ij} & = & 4 \pi \int_{0}^{\infty} W_{ErF,ij} r_{ij}^{2} dr_{ij} \nonumber \\ & = & 4 \pi \int_{0}^{\infty} \frac{1}{2 \pi^{3/2} h r_{ij}^{2}} e^{- r_{ij}^{2}/h^{2}} r_{ij}^{2} dr_{ij} \nonumber \\ & = & 4 \pi \int_{0}^{\infty} \frac{1}{2 \pi^{3/2} h} e^{- r_{ij}^{2}/h^{2}} dr_{ij} \nonumber \\ & = & \frac{2}{\sqrt{\pi}} \int_{0}^{\infty} e^{- q_{ij}^{2}} dq_{ij}, \ q_{ij} = r_{ij}/h.
\end{eqnarray}

  Hence, $\int W_{ErF,ij} d^{3} r_{ij} = 1$.
  
  Also $\int W_{3S,ij} d^{3} r_{ij} = 1$, as well as $\int W_{G,ij} d^{3} r_{ij} = 1$, this last, considering the well known properties of Gaussian integrals: $I_{n} = \int_{0}^{\infty} t^{n} e^{- \xi t^{2}} dt$, and in particular $I_{2} = \int_{0}^{\infty} t^{2} e^{- t^{2}} dt = \pi^{1/2}/4$.

  Fig. 1 displays, $W_{3S,ij}$ $W_{G,ij}$ and $W_{ErF,ij}$ as a function of $q_{ij} =  r_{ij}/h$. $4 \pi q_{ij}^{2} W_{3S,ij}$, $4 \pi q_{ij}^{2} W_{G,ij}$ and $4 \pi q_{ij}^{2} W_{ErF,ij}$, as well as $4 \pi q_{ij}^{2} \nabla W_{3S,ij}$, $4 \pi q_{ij}^{2} \nabla W_{G,ij}$ and $4 \pi q_{ij}^{2} \nabla W_{ErF,ij}$ are significant for 3D integrations. Fig. 1 displays the much better GASPHER interpolation capabilities, with respect to the current SPH or ASPH techniques using other Kernels, not only because 3D interpolations are more weighted toward $r_{ij} \rightarrow 0$, but also because $- \nabla W_{ErF,ij} \rightarrow \infty$ as it should be, avoiding the well known "particle pairing instability" effect, affecting the other two behaviours ($\nabla W_{ij}$ displays a minimum for $r_{ij}/h \approx 1$). In the conversion from mathematical integrals to computational summations in 3D, the role of $4 \pi r_{ij}^{2} dr_{ij}$ is equivalent to $n_{i}^{-1}$. Thus, wherever $n_{i}^{-1} \gg h^{3}$, and spatial gradients exist, the effectiveness of the adopted interpolation Kernel comes out. In the resolution of the Euler or of the Navier Stokes equations, spatial derivatives have to be calculated. In the calculation of $\nabla p/\rho$ in the momentum equation, two particles cannot coincide because the pressure force is physically infinite. Moreover, also for the $\nabla \cdot \bmath{v}$ in the energy equation or in the continuity equation, this non physical case should be carefully avoided because no velocity divergence can exist if particle mutual separation is zero. Summing up, both indexes $i \neq j$ and $r_{ij} > 0$.  In the unrealistic case of $r_{ij} = 0$, spatial derivatives to compute gradients or divergences can be bypassed because unphysical. In the case of a very short particle mutual separation, natural computational difficulties can arise only for a very small particle separation. This is unavoidable when a very high compression characterize the fluid, because calculated pressure and individual pressure forces are always naturally very high. However, in particular for accretion or collapse processes, the particle merging in a new particle, created at the centre of mass, conserving mass, energy and momentum could be the best solution. This is a useful physical expedient, also used in ASPH, whenever a strong gas compression occurs. It avoids a too short explicit time step calculation, according to the well known Friedrich-Courant-Lewy. In particular, for ASPH technique only, it also avoids any artificial viscosity inadequacy in handling shocks. Such an allowed expedient could be correctly also used in GASPHER in such conditions.

%-------------------------------------------------- FIG 1 START
\begin{figure}
%\hspace{25 mm}
\includegraphics[width=8cm,height=8cm]{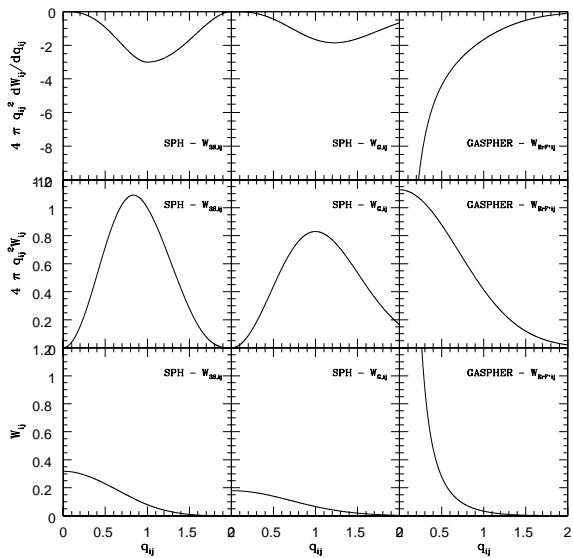}
\caption{Radial plots of SPH Kernels $W_{3S,ij}$ (eq. 7), $W_{G,ij}$ (eq. 6), as well as of GASPHER $W_{ErF,ij}$ (eq. 8). $q_{ij} = r_{ij}/h$. $4 \pi q_{ij}^{2} W_{3S,ij}$, $4 \pi q_{ij}^{2} W_{G,ij}$, $4 \pi q_{ij}^{2} W_{ErF,ij}$, as well as the radial derivatives times $4 \pi q_{ij}^{2}$ both useful for 3D calculations are also reported.}
\end{figure}
%-------------------------------------------------- FIG 1 END

  We pay attention that in 3D interpolations, it is not the role of the Kernel $W$ that it is important. Instead, it is the $W r^{2}$ that is to be taken into account, as Fig. 1 clearly displays. Hence, on $ith$ particle, when $r_{ij} \rightarrow 0$, $W r^{2}$ converges toward a finite value. If this is the explanation regarding the continuum limit, in the spatial discretization, this role is carried out by the particle density $\rho_{j}$ which, in the SPH formulation $A_{i} = \sum_{i} A_{j} W_{ij}/n_{j} = \sum_{i} m_{j} A_{j} W_{ij}/\rho_{j}$, divides $W_{ij}$. Only when $r_{ij} \rightarrow 0$ GASPHER formulation becomes: $A_{i} = \sum_{i} A_{j} W_{ij}/n_{j} \rightarrow 4 \pi A_{j} W_{ij} r_{ij}^{2} h^{3}$. On the other hand, whenever $r_{ij} \rightarrow 0$ the concept of dimension is meaningless. This implies that, if $r_{ij} \rightarrow 0$, and especially if $r_{ij} = 0$, the 1D formulation of Kernel can also be taken into account to simplify computational complications in some selected cases, whenever the 2D or the 3D fluid kinematics flows along one selected direction.

\subsection{Advantages of GASPHER}

This Kernel choice resolves the problem of neighbours inadequacy, as well as the problem of the SPH and ASPH "particle pairing instability" for $r_{ij}/h < 1$ due to the fact that when $r_{ij} \rightarrow 0$, $- \nabla p$ does not become infinite.

  For practical reasons in computational resources, even a limitation to several $h$ of the order of $lh$ with $l \sim 4 \div 10$ could be considered with very small modifications in results, keeping constant the resolving power $h$ of all particles. In fact, theoretically considering a homogeneous and isotropic 3D particle distribution, if $4 \pi nh^{3}/3$ (where $n$ is the particle concentration) is the number of neighbours closer than $h$ for each ith particle, it increases up to $4 \pi n(lh)^{3}/3$ i.e. up to $\sim 64 \div 10^{3}$ times. Alternatively, neighbours can be limited to a selected number ($40$ in our 3D models). In both cases, a very small modifications in results, neglecting further interpolating particles, is made because the most important neighbours in the interpolation are the closest ones. In this case, if neighbours are a large number, due to a very high particle concentration, it is easily possible to merge more particles in a single new particle, created at the centre of mass, conserving mass, energy and momentum. So doing, the ASPH's risk to decrease the spatial smoothing resolution length to values involving an ineffective artificial viscosity behaviour, as well as the danger to get a too small computed time step in the Courant-Friedrich-Lewy condition when $h \rightarrow 0$, are avoided. A fixed number of neighbours can be a serious risk by limiting the interaction to $30 \div 40$ neighbours only the inner "flat part" of the Kernel contributes to SPH sums. A large contribution from other particles outside this "flat part" would be wrongly neglected. In GASPHER this does not occur because the Kernel slope is not "flat" for $r_{ij} \rightarrow 0$, instead $\nabla W_{ErF,ij} \rightarrow - \infty$. Of course, also ASPH techniques try to avoid the unpleasant "particle pairing instability". However, the particle resolving power $h$ cannot decrease too much in regions of very high particle concentrations otherwise the artificial dissipation due to the artificial viscosity does not work well. Moreover, at the same time, the time step explicitly computed according to the Friedrich-Courant-Lewy condition becomes too short if $h \rightarrow 0$.

  The possibility of adopting a numerical SPH code, including the physical viscosity  \citep{b10,b11}, considering \citet{b28,b19,b20,b56} results, makes us able to answer the problem whether ASPH's and/or GASPHER methods are reliable in improving fluid dynamics compared to the original SPH, where the smoothing length $h$ is constant. Although some authors \citep{b12,b52,b48} adopted Gaussian Kernels, their methods belong to the ASPH numerical schemes where a spatially and temporarily variable smoothing length $h$ is adopted. Although many efforts try to conciliate a reliable adaptive interpolation technique with computational resources, ASPH methods are unsatisfactory in describing a correct gas dynamics because hidden numerical errors exist inside an adaptive interpolation, better revealed in a viscous transport process inside a definite potential well. All ASPH's difficulties in handling the artificial viscosity dominant role in subsonic and/or expanding regimes, as discussed before, are prevented in GASPHER by the fact that the particle resolving power $h$ is constant and equal to HWHM of spatially unlimited Gaussian Kernels. GASPHER technique limits the problem of particle disorder in computing particle $\nabla \cdot \bmath{v}$, as discussed in \citet{b16} and in \citet{b40} as far as shear flows are concerned because, even considering disordered flows, particle disorder is tamed by GASPHER extended interacting particle domains. In fact, the longer the particle interpolation range, the better the computational result, without any modification of particle resolving power $h$.

  Finally, an adequate fixed smoothing resolution length $h$ allows us to resolve gas turbulence within the confined integration domain even in low compressibility regimes. In non viscous conditions the local Reynolds number $Re = L v/\nu \sim 10^{2} v/c_{s}$, considering $L = 0.5$, $\nu \approx c_{s} h$ \citep{b34} and, more stressing, $Re \sim 10^{3} v/c_{s}$ considering $\nu \approx 0.1 c_{s} h$ \citep{b44,b47}, because of $\alpha_{SPH} = 1$. Being the whole disc structure typically supersonic, even for $\gamma = 5/3$, $Re > 10^{3}$. Hence, a moderate turbulence is effective in non viscous $\gamma$ conditions, where gas collisions are relevant. Instead in viscous conditions, in the Shakura and Sunyaev formulations, no turbulence is recorded.

  If an adaptive method is adopted in low compressibility conditions, the increasing of the smoothing resolution length $h$, up to an order of magnitude, prevents any turbulence resolution in an accretion disc, even for supersonic regimes. The evaluation of the minimum linear dimensions of the integration domain, able to solve turbulence adopting an $\alpha_{SS}$ parameter of the order of $0.1-0.5$, gives a value of the order $D \geq 10^{-2} \div 10^{-1}$ in order to get a Reynolds number $Re \geq 10^{2}$, the smaller value is for supersonic regimes. The integration domain (the length of the primary's potential well) of the order of $0.5$. Therefore, how to handle an adaptive SPH with the problem of solving the turbulence is a real difficulty, and the adopted fixed $h$ is correct in order to solve this problem. Larger (and adaptive) $h$ values are in open conflict with $D \geq 10^{-2} \div 10^{-1}$ in order to solve turbulence.

  These conclusions on turbulence in accretion discs having free edge boundaries are not those concerning the concept of turbulence wherever fixed static boundaried are considered. Whenever particles move within a confined box, both SPH and ASPH results are traditionally correct in so far as $r_{ij}/h$ is not too small. In this case the problem regards the particle chaotic collisions in a close environment where the particle mean free path is less than two or three times the particle smoothing resolution length.

%
%__________________________________________________________________
%
%

\section{GASPHER disc simulations in CB: results and discussion}

  Looking at our SPH results in a physically viscous low compressibility regime  \citep{b28,b19,b20,b56} as a reference, where the particle smoothing length is constant and a typical cubic spline function as a smoothing function have been assumed, we systematically perform a series of GASPHER simulations with the aim of getting a physically viscous well-bound accretion disc in a close binary. We show that such transport phenomenology, in a low compressibility regime, is significant in deciding the reliability of the adopted Kernel formulation for SPH fluid dynamics simulations, especially whenever free edge boundary conditions must be taken into account. 

\subsection{Parameters and boundary conditions}

  The characteristics of the binary system are determined by the masses of the two companion stars and their separation. We chose to model a system in which the mass $M_{1}$ of the primary compact star and the mass $M_{2}$ of the secondary normal star are equal to $1 M_{\odot}$ and their mutual separation is $d_{12} = 10^{6} \ Km$. The primary's potential well is totally empty at the beginning of each simulation at time $T = 0$. The injection gas velocity at L1 is fixed to $v_{inj} \simeq 130 \ Km \ s^{-1}$ while the injection gas temperature at L1 is fixed to $T_{\circ} = 10^{4} \ K$, taking into account, as a first approximation, the radiative heating of the secondary surface due to lightening of the disc. Gas compressibility is fixed by the adiabatic index $\gamma = 5/3$. Supersonic kinematic conditions at L1 are discussed in  \citet{b28,b56}, especially when active phases of CB's are considered. However, results of this paper are to be considered as a useful test to check whether disc structures (viscous and non) show the expected behaviour. The reference frame is that centred on the primary compact star and corotating, whose rotational period, normalized to $2 \pi$, coincide with the orbital period of the binary system. This explain why in the momentum equation (eq. 2), we also include the Coriolis and the centrifugal accelerations.

   In our models the unknowns are: pressure, density, temperature, velocity, therefore we solve the continuity, momentum, energy, and state (perfect gas) equations. In order to make our equations dimensionless, we adopt the following normalization factors: $M = M_{1} + M_{2}$ for masses, $d_{12} = 10^{11} \ cm$ for lengths, $v_{\circ} = (G(M_{1} + M_{2})/d_{12})^{1/2}$ for speeds, so that the orbital period is normalized to $2 \pi$, $\rho_{\circ} = 10^{-9} \ g \ cm^{-3}$ for the density, $p_{\circ} = \rho_{\circ} v_{\circ}^{2} \ dyn \ cm^{-2}$ for pressure, $v_{\circ}^{2}$ for thermal energy per unit mass and $T_{\circ} = (\gamma -1) v_{\circ}^{2} \ m_{p} \ K_{B}^{-1}$ for temperature, where $m_{p}$ is the proton mass and $K_{B}$ is the Boltzman constant. The adopted Kernel resolving power in the GASPHER modelling is $h = 5 \cdot 10^{-3}$. The geometric domain, including moving disc particles, is a sphere of radius $0.6$, centred on the primary. The rotating reference frame is centred on the compact primary and its rotational period equals the orbital one. We simulated the physical conditions at the inner and at the outer edges as follows:

a) inner edge: \\
the free inflow condition is realized by eliminating particles flowing inside the sphere of radius $2 \cdot 10^{-2}$, centred on the primary. Although disc structure and dynamics are altered near the inner edge, these alterations are relatively small because they are counterbalanced by a high particle concentration close to the inner edge in supersonic injection models.

b) outer edge: \\
the injection of "new" particles from L1 towards the interior of the primary Roche Lobe is simulated by generating them in fixed points, called "injectors", symmetrically placed within an angle having L1 as a vertex and an aperture of $\sim 57^{\circ}$. Normally, as adopted since our first paper on SPH accretion disc in CB \citep{b34}, the radial elongation of the whole ensemble of injectors is $\sim 10 h$. The initial injection particle velocity is radial with respect to L1. In order to simulate a constant and smooth gas injection, a "new" particle is generated in the injectors whenever "old" particles leave an injector free, inside a small sphere with radius $h_{min}$, centred on the injector itself. Particle masses are determined by the assumed local density at the inner Lagrangian point L1: $\rho_{L1} = 10^{-9} g \ cm^{-3}$ (as typical stellar atmospheric value for the secondary star), equal to $m = \rho_{L1} (h d_{12})^{3}/(M_{1} + M_{2})$.

  The formulation adopted for the 3D SPH viscous accretion disc models is the well-known $\alpha_{SS}$ \citet{b50,b70} and \citep{b51} parametrization: $\nu_{SS} = \alpha_{SS} c_{s} H$, where $c_{s}$ is the sound velocity, $0 \leq \alpha_{SS} \leq 1$ and $H = r_{xy}c_{s}/(M_{1}/r_{xy})^{1/2}$ is a dimensionless estimate of the Standard disc thickness, where $r_{xy} = (X_{i}^{2} + Y_{i}^{2})^{1/2}$ is the cylindrical radial coordinate of the ith particle. In this paper we adopt $\alpha_{SS} = 1$ to point out evident differences in disc structure and dynamics between our disc models.

\subsection{General results}

  We carried out our low compressibility ($\gamma = 5/3$) simulations until we achieved fully stationary configurations. This means that particles injected into the primary potential well (which is not deep, according to the primary small mass) are statistically balanced by particles accreted onto the primary and by particles ejected from the outer disc edge.

%-------------------------------------------------- FIG 2 START
\begin{figure}
%\hspace{25 mm}
\includegraphics[width=8cm,height=8cm]{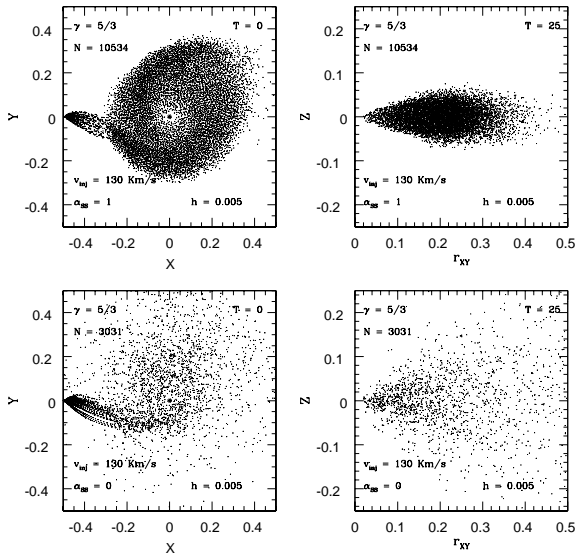}
\caption{XY plots and $r_{XY}Z$ plots for both the inviscid ($\alpha_{SS} = 0$) and the viscous ($\alpha_{SS} = 1$) disc model. The final time $T$ and the total particle number $N$, as well as the injection velocity from the inner Lagrangian point L1 and $h$, are also reported.}
\end{figure}
%-------------------------------------------------- FIG 2 END

  The orders of magnitude of the mass transfer injection rate from L1: $\dot{M}_{inj}$, the accretion rate $\dot{M}_{acc}$ and the ejection rate $\dot{M}_{eje}$ are $\approx 10^{17}$, $\approx 5.5 \cdot 10^{16}$ and $\approx 4.5 \cdot 10^{16}$, for the non viscous model and $\approx 8.0 \cdot 10^{16}$, $\approx 7.9 \cdot 10^{16}$ and $\approx \cdot 10^{15}$ for the viscous model, respectively. $\rho_{\circ} h^{3} d_{12}^{2} v_{\circ}$, is the conversion factor from particle/time to $g \ s^{-1}$. Such values (also adopted in \citet{b28,b19,b20,b56}) are representative of active phases of CB whenever either the restricted problem of three bodies in terms of the Jacobi constant or the Bernoulli's theorem are taken into account during such phases \citep{b14}, considering the conservation of the flux momentum in the crossing of L1 from the two Roche lobes.

  Fig. 2 displays XY plots of both the physically inviscid and viscous disc models ($\alpha_{SS} = 0$ and $\alpha_{SS} = 1$). $N$ represents the total number of particles in each model. In classical SPH, no well-bound structures with a definite disc's outer edge come out (Molteni et al. 1991; Lanzafame et al. 1992, for $\gamma = 1.1$ and $\gamma = 1.2$). The inviscid GASPHER disc model shows a higher particle concentration at the disc's inner edge, close to the primary star. Instead, a well-defined structure comes out in the viscous disc model in stationary conditions.

  Fig. 2 also displays the $r_{XY}Z$ plots obtained by folding all disc bulk particles onto a plane containing the $Z$ axis and being perpendicular to the XY orbital plane. An evident latitudinal spread appears for the inviscid model. Computed latitudinal angular spread is $\approx 60^{\circ}$ for inviscid disc model. This results compare to that obtained in \citet{b34} and in \citep{b23}, as far as the non viscous model, and to that obtained in \citet{b28,b56} as far as the viscous model, are concerned. As for the viscous model, the latitudinal spread is $\approx 24^{\circ}$.

  Fig. 2 clearly displays the coming out of spiral patterns in the XY plot of the viscous model. These particular structures did not come out in SPH \citet{b28} results, where both the same supersonic injection conditions from L1, and the same stellar masses, as well as the same low gas compressibility were adopted in viscid $\alpha_{SS} = 1$ conditions. However, an exhaustive literature \citep{b59,b60,b61,b58,b57,b26,b27}  exists, showing which conditions favour the development of such structures (e.g. tidal torques, external and/or outer edge perturbations). In particular, \citet{b26,b27}  showed that high angular momentum injection condition from L1 produces these patterns. This beyond doubt shows the better effectiveness of GASPHER Kernel choice (33) compared to the common cubic spline SPH Kernel analytical formulation.

  The comparison with \citet{b28,b56} results ensures us that GASPHER technique performs not only correct calculations, but also that particles at disc's outer edge are not isolated. Moreover, the full radial transport cannot be affected by any "particle pairing instability" because the Kernel formulation (33) prevents such unpleasant inconsistency in the disc bulk.

  Low compressibility gas loss effects affect the non viscous GASPHER disc surfaces and outer edge. The same result were obtained in \citet{b34}, as well as in \citet{b23}, working in SPH and adopting two different spatial smoothing resolution lengths and sonic injection transfer conditions from L1. Supersonic injection conditions from L1 are now taken into account, as described in \citet{b56} for active phases of CB. Therefore, non viscous gas loss effects from disc's outer edge and surfaces are {\it a fortiori} correctly expected, taking into account of the higher injection mechanical energy from L1. At the same time, even though the low density non viscous disc structure is statistically rarefied, no neighbour inadequacy affect GASPHER interpolation.

  The low total number of particles within the primary's potential well ($\sim 3000$) in non viscous conditions is due to the absence of any physical viscosity able to keep bound particles against pressure forces responsible of particle removal from the disc outer free edge for $\gamma = 5/3$ whenever low mass CB's are considered. This result is well known \citep{b34,b24,b20,b56}. This particular is not trivial because the bound of the edge of the computational domain prevents any particle removal allowing to get the wished particle concentration in spite of the effective particle repulsion for high $\gamma$ values.

%
%__________________________________________________________________
%
%

\subsection{Accuracy}

  SPH free surface flows were developed by \citet{b37}, with the aim of solving the unpleasant problem of FE layer in SPH techniques, mainly to simulate breaking waves, but at relatively low resolution. A reduction in noise, with smooth-free surfaces and regular particle distribution, was obtained by \citet{b4} and \citet{b5}, developing SPH models where first order completeness was enforced, that is that first order polynomials are exactly reproduced. Error estimates in a SPH interpolant are evaluated in \citet{b35,b36}. However in this paper, the lack of completeness of SPH interpolants is not taken into account. A formulation for the total error, determining how simulation parameters should be chosen and taking into account of the order of completeness is still not written in the literature. \citet{b5} adopted modified Kernel gradients into the classical SPH equations. However, the hidden problem with this approach is that modified Kernels no longer have the property that spatial gradients with respect to their two position arguments are exactly opposite between two contact particles. This Kernel property is essential in SPH equations. \citet{b55} showed "an expression for the error in an SPH estimate, accounting for completeness, an expression that applies to SPH generally", paying attention to the conservation principles. They found that a common method, enforcing completeness, violates the conservation principle of Kernel spatial gradients must be opposite between two contact particles. They also showed some examples of discretization errors: numerical boundary layer errors. Errors for a SPH summation interpolant are functions of both particle distribution and particle smoothing length \citep{b35,b55}. In an exact formulation, such errors are described by both volume and surface integrals of both neighbour particle distribution and their smoothing resolution length $h$. Therefore, in FE layer conditions, not only relevant errors in interpolations, but also unnatural pressure gradients in FE conditions at the edge of the computational domain occur in ASPH.

%-------------------------------------------------- FIG 3 START
\begin{figure}
%\hspace{25 mm}
\includegraphics[width=8cm,height=8cm]{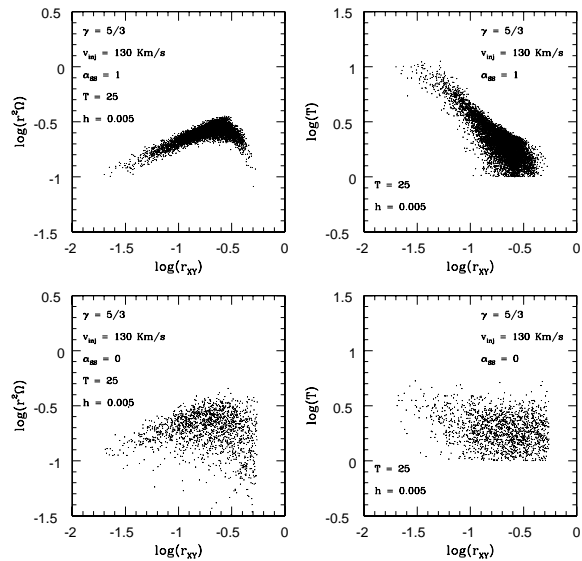}
\caption{Logarithmic plots of specific angular momentum radial distributions and of temperature for both the inviscid models ($\alpha_{SS} = 0$), and the viscous models ($\alpha_{SS} = 1$). The final time $T$ and the total particle number $N$, as well as the injection velocity from the inner Lagrangian point L1 and $h$, are also reported.}
\end{figure}
%-------------------------------------------------- FIG 3 END

  A reasonable SPH accuracy is related to the number of space neighbours of each SPH particle. As a free Lagrangian numerical method, classical SPH methods, are free from errors as far as momentum and angular momentum are concerned. Instead, errors can occur as to energy as for ASPH variants. In particular, it is remarkable the evaluation of the energy error propagator (EEPR) $(\Delta E/dt)/E$, computed for each particle, to have the correct idea of temporal propagation of energy errors. If SPH-like methods involve a systematic error in energy [$\Delta E/E] |_{err}$ of a few percent, this error progressively increases in time as $\int [(dE/dt)/E]|_{err} dt$. This means that, if a long time is necessary to achieve a fully stationary configuration, errors in energy conservation could be significant. The evaluation of the GASPHER EEPR for inviscid disc model is $\approx 3.02 \cdot 10^{-6}$. Instead, the GASPHER EEPR for viscous disc model is $\approx 1.85 \cdot 10^{-6}$. Being errors in energy of this order of magnitude we do not usually allow to distinguish if numerical simulations correspond to a fluid physical behaviour. Thus, the numerical error in energy on particle over-expansion/over-compression is not dominant step by step. Unfortunately, it accumulates in time. This implies that numerical simulations, limited only to explosive or collapse short time tests, would not be reliable in testing ASPH codes. Therefore, once more, this conclusion strengthens numerical tests and simulations based on a transport mechanism.

\subsection{The role of physical viscosity}

  Physical viscosity naturally works where the particle mutual velocities (and separations) change in time, namely when a mutual acceleration exists, contrasting gas dynamics (rarefaction or compression) and converting kinetic energy in thermal energy. Such a mechanism clearly supports the development of well-bound accretion discs inside the primary potential well, in spite of the low compressibility, at least for $\alpha_{SS} = 1$ both in classical SPH and in GASPHER approach.

  We want to point out that adopting $\alpha_{SS} = 1$ does help emphasize differences in disc structure and dynamics compared to the physically inviscid model. However, values of $\alpha_{SS}$ smaller than the unity may be more realistic according to some thin disc analytical models \citep{b49,b29}. We recall that our physical viscosity is only a shear viscosity. For the sake of simplicity, no bulk viscosity has been considered, as explicitly mentioned in the paper. In fact, a value $\alpha_{SS} = 1$ for the bulk viscosity should be too high. Fig. 3 displays, in a logarithmic scale, the angular momentum and temperature radial distribution, for all models. Such radial distributions for the GASPHER viscous model are very close to that of the Standard model $r^{2} \Omega \propto r^{1/2}$ and $T \propto r^{-3/4}$. This can be explained considering that, in stationary conditions, an accretion disc redistributes the angular momentum injected at the outer edge into the disc bulk, according to outer edge boundary conditions only, as already shown in \citet{b1} and \citet{b24}. Physical viscosity plays a role in regions where particle velocity gradients are significant. This means that physical viscosity plays a relevant role mainly in the radial transport, while it has scarce influence on the tangential dynamics. A strong difference appears when looking at the temperature radial distribution. In fact, the heating effect of the physical viscosity is particularly evident in the disc's inner zones. We recall that the disc itself is in an equilibrium stationary state where the heated particles are directly accreted towards the primary. This as far as particle advection is concerned. As for conduction, although it is much less important, notice that the temperature decreases towards the exterior, thus dispersing heat outside. However, discs could also radiate energy. In disc models without explicit inclusion of radiative terms in the energy equation (almost all models, since, this inclusion complicates things considerably), the effect of radiative cooling is better simulated with $\gamma$'s less than $5/3$.

%-------------------------------------------------- FIG 4 START
\begin{figure}
%\hspace{25 mm}
\includegraphics[width=8cm,height=8cm]{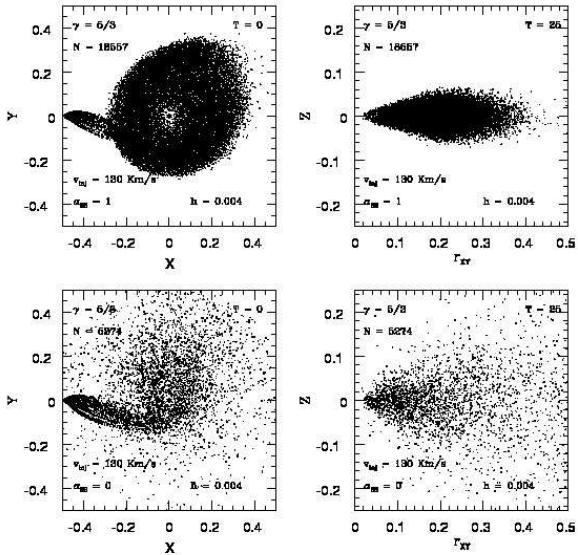}
\caption{XY plots and $r_{XY}Z$ plots for the inviscid and the viscous ($\alpha_{SS} = 0$ and $\alpha_{SS} = 1$) disc models for a different particle smoothing resolution lengths $h = 4 \cdot 10^{-3}$. The final time $T$ and the total particle number $N$, as well as the injection velocity from the inner Lagrangian point L1, are also reported.}
\end{figure}
%-------------------------------------------------- FIG 4 END

%-------------------------------------------------- FIG 5 START
\begin{figure}
%\hspace{25 mm}
\includegraphics[width=8cm,height=8cm]{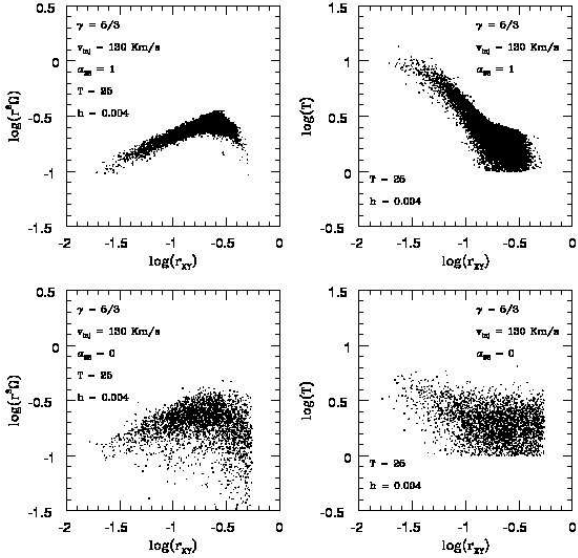}
\caption{Logarithmic plots of the radial distribution of the specific angular momentum and of temperature for both viscous and non viscous accretion discs. $\alpha_{SS}$, the particle smoothing resolution lengths $h$, as well as the final time $T$, the total particle number $N$, and the injection velocity from the inner Lagrangian point L1, are also reported.}
\end{figure}
%-------------------------------------------------- FIG 5 END

\section{Influence of the GASPHER smoothing resolution length on disc structure: does the spatial resolving power affect general result?}

  To study how FE fluid dynamics is affected by the initial smoothing resolution length choice, we performed two more simulations (non viscous and viscous) adopting a smaller smoothing resolution length: $h = 4 \cdot 10^{-3}$, improving spatial resolution since particle injection. Taking into account of injection condition from our previous simulations for $h_{i, L1} = 5 \cdot 10^{-3}$, particle masses are scaled, conserving the same mass density from L1, according to the ratio of particle volumes: $m_{i, h} = m_{i, h=0.005} (h_{i, L1}/5 \cdot 10^{-3})^{3}$. Thus, the mass transfer rate from L1 is self-consistent and automatically comparable to that relative to simulations with $h_{i, L1} = 5 \cdot 10^{-3}$, without any variation of the injection velocity $v_{inj} \simeq 130 \ Km \ s^{-1}$. To do this, it is necessary to recalculate newly the total number of injectors, by adopting the simple scale law: $N_{inj} = N_{inj, h=0.005} (5 \cdot 10^{-3}/h_{i, L1})^{3}$. Hence, according to these simple scaling laws, we keep injection conditions comparable both for the initial density and for the mass transfer rate at L1.

  Results of such further simulations are displayed in Figg. 4 and 5, where plots of such $\gamma = 5/3$ 3D GASPHER simulations are displayed, free of any difficulty on the sufficient number of neighbour particles, as explained before. The total number of disc particles shows a monotonic increase by decreasing $h$. Thus, both disc density is comparable in both disc models as well as the mass of the simulated discs being $N h_{i}^{3}$, as well as $N m_{i} \approx$ constant, within statistical fluctuations.

  In the non viscous regime the larger number of disc particles are still affected by a gas chaotic collisional component on top of the spiral disc's kinematics. Moreover, the Reynolds number increases because of the reduction of the particle smoothing resolution length, from $h = 5 \cdot 10^{-3}$ to $h = 4 \cdot 10^{-3}$. Instead, whichever is the GASPHER adopted particle smoothing resolution length $h$, in a viscous regime, both the radial transport of mass and angular momentum, as well as the radial temperature profile, are not sensitive to any adopted particle resolving power as Figg. 2 to 5 clearly display. Their radial behaviour is strictly comparable to that of the typical standard disc, whose specific angular momentum $r^{2} \Omega \propto r^{1/2}$ and whose mean temperature $T \propto r^{-3/4}$. Hence, this result is a further confirming check that GASPHER result, in their general aspect, as far as the radial transport and thermal properties, are not strongly dependent on the assumed spatial resolution. Moreover, the local physical properties are clearly comparable with each other, being the particle spatial resolution in the graphs different, but not their physical values, that is denser (more rarefied), lighter (heavier) particles to get the same density, as an example.

\section{Concluding remarks}

  From the astrophysical point of view, our results show that in GASPHER modelling, where particle interpolation radial extension is conceptually unlimited - although particle smoothing length $h$ is spatially and permanently constant - solve the problem of neighbours inadequacy. Moreover, physical viscosity supports the development of a well-bound accretion disc in the primary potential well, even in the case of a low compressibility gas dynamics. Such results, also shown in \citet{b28,b19,b20,b56}, mean, once more, that the initial angular momentum injection conditions at the disc's outer edge are responsible for the disc tangential dynamics, while viscosity is mainly responsible for the thermodynamical disc properties, even for low compressibility disc models ($\gamma > 1.1$, here $\gamma = 5/3$) when gas loss effects are physically expected according to the low compressibility gas dynamics and to the low stellar mass of the central accretor. Moreover, in GASPHER viscous fluid dynamics, further details of the flow are revealed (e. g. the coming out of spiral patterns in disc structures).

  From the numerical point of view, reliable results are reproduced in a GASPHER, despite FE conditions are adopted. Without considering the injected particle stream, such simulations could also be considered as accretion and transport general tests within a gravitational potential well. Typical tests as far as non viscous 1D shock tube show that GASPHER technique produce results in a very good comparison with analytical ones, having the advantage to solve the FE difficulties without any "particle pairing instability". Simulation, carried out in low compressibility and in high viscosity conditions, to stress out results, is significant to understand the quality of numerical code. The transformation of SPH codes in a GASPHER code, without further numerical efforts, seems likely to be an interesting future challenge. As far as the computational cpu time is concerned, there is not conceptually any disadvantage in such transformation, if particle neighbours are fixed (e.g. $30$ or $50$) for each particle, by the introduction of a boundaries counter/limiter because the number of particle neighbours rules the computational cpu time.

  The necessity to perform better SPH numerical interpolations on contact surfaces, or at FE layers, recently inspired authors to develop SPH-derived techniques to achieve a higher accuracy. An SPH dynamic refinement has recently been developed by \citet{b9} to calculate boundary contact forces in fluid flow problems through boundary particle splitting. Such a technique could also be very interesting and competitive in solving FE problems. However, this is beyond the scope of this paper.

  We conclude that although high compressibility inviscid results among different schemes could compare with each other especially if constraints are imposed on boundaries of the computational domain, differences arise either if FE and/or if viscous flows are involved. In such conditions, GASPHER technique shows a regular behaviour and better conserve the total energy, as well as reduces the influence of the artificial viscosity for non viscous ideal shear flows free of any gas compression (see Appendix). Computational cpu time is mainly governed by the number of neighbour particles for each particle. Therefore, no disadvantages arise, in principle, in adopting a GASPHER code with respect to an ASPH code if the neighbour particle statistical number is the same.

%
%__________________________________________________________________
%
%

\appendix

\section{SPH formulation of both physically inviscid and viscous perfect gas 
hydrodynamics}

\subsection{SPH and ASPH (in adaptive smoothing length $h$) techniques}

  The SPH method is a Lagrangian scheme that discretizes the fluid into moving interacting and interpolating domains called "particles". All particles move according to pressure and body forces. The method makes use of a Kernel $W$ useful to interpolate a physical quantity $A(\bmath{r})$ related to a gas particle at position $\bmath{r}$ according to:

\begin{equation}
A(\bmath{r}) = \int_{D} A(\bmath{r}') W(\bmath{r}, \bmath{r}', h) d \bmath{r}'
\end{equation}

$W(\bmath{r}, \bmath{r}', h)$, the interpolation Kernel, is a continuous function - or two connecting continuous functions whose derivatives are continuous even at the connecting point - defined in the spatial range $2h$, whose limit for $h \rightarrow 0$ is the Dirac delta distribution function. All physical quantities are described as extensive properties smoothly distributed in space and computed by interpolation at $\bmath{r}$. In SPH terms we write:

\begin{equation}
A_{i} = \sum_{j=1}^{N} \frac{A_{j}}{n_{j}} W(\bmath{r}_{i}, \bmath{r}_{j}, h) = \sum_{j=1}^{N} \frac{A_{j}}{n_{j}} W_{ij}
\end{equation}

where the sum is extended to all particles included within the domain $D$, $n_{j} = \rho_{j}/m_{j}$ is the number density relative to the jth particle. $W(\bmath{r}_{i}, \bmath{r}_{j}, h) \leq 1$ is the adopted interpolation Kernel whose value is determined by the relative distance between particles $i$ and $j$.

  In SPH conversion of mathematical equations (eq. 1 to eq. 4) there are two principles embedded. Each SPH particle is an extended, spherically symmetric domain where any physical quantity $f$ has a density profile $f W(\bmath{r}_{i}, \bmath{r}_{j}, h) \equiv f W(|\bmath{r}_{i} - \bmath{r}_{j}|, h) = f W(|\bmath{r}_{ij}|,h)$. Besides, the fluid quantity $f$ at the position of each SPH particle could be interpreted by filtering the particle data for $f(\bmath{r})$ with a single windowing function whose width is $h$. So doing, fluid data are considered isotropically smoothed all around each particle along a length scale $h$. Therefore, according to such two concepts, the SPH value of the physical quantity $f$ is both the overlapping of extended profiles of all particles and the overlapping of the closest smooth density profiles of $f$. This means that the compactness of the Kernel shape gives the principal contribution to the interpolation summation to each particle by itself and by its closest neighbours. In both approaches the mass is globally conserved because the total particle number is conserved.

  In SPH formalism, equations (2) and (3) take the form:
  
\begin{eqnarray}
\frac{d \bmath{v}_{i}}{dt} & = & - \sum_{j=1}^{N} m_{j} 
\left( \frac{p_{i}^{\ast}}{\rho_{i}^{2}} + \frac{p_{j}^{\ast}}{\rho_{j}^{2}} \right) \nabla_{i} W_{ij} + \bmath{g}_{i} + \nonumber \\
& & \sum_{j=1}^{N} m_{j} \left( \frac{\eta_{vi} \bmath{\sigma}_{i}}{\rho_{i}^{2}} + \frac{\eta_{vj} \bmath{\sigma}_{j}}{\rho_{j}^{2}} \right) \cdot \nabla_{i} W_{ij} \\
\frac{d}{dt} E_{i} & = & - \sum_{j=1}^{N} m_{j} \left( \frac{p_{i}^{\ast} \bmath{v}_{i}}{\rho_{i}^{2}} + \frac{p_{j}^{\ast} \bmath{v}_{j}}{\rho_{j}^{2}}\right) \cdot \nabla_{i} W_{ij} + \bmath{g}_{i} \cdot \bmath{v}_{i} + \nonumber \\
& & \sum_{j=1}^{N} m_{j} \left( \eta_{vi} \frac{\bmath{\sigma}_{i} \cdot \bmath{v}_{i}}{\rho_{i}^{2}} + \eta_{vj} \frac{\bmath{\sigma}_{j} \cdot \bmath{v}_{j}}{\rho_{j}^{2}} \right) \cdot \nabla_{i} W_{ij}
\end{eqnarray}

where $\bmath{g}_{i} = - 2 \bmath{\omega} \times \bmath{v}_{i} + \bmath{\omega} \times (\bmath{\omega} \times \bmath{r}_{i}) - \nabla \Phi_{grav, i}$, $\bmath{v}_{ij} = \bmath{v}_{i} - \bmath{v}_{j}$, $m_{j}$ is the mass of jth particle and $p_{i}^{\ast} = p_{i} +$ {\it artificial pressure term}. $E_{i} = (\epsilon_{i} + \frac{1}{2} v_{i}^{2})$.  The viscous stress tensor $\tau_{\alpha \beta}$ includes the positive first and second viscosity coefficients $\eta_{v}$ and $\zeta_{v}$ which are velocity independent and describe shear and tangential viscosity stresses ($\eta_{v}$), and compressibility stresses ($\zeta_{v}$):

\begin{equation}
\tau_{\alpha \beta} = \eta_{v} \sigma_{\alpha \beta} + \zeta_{v} \nabla 
\cdot \bmath{v}
\end{equation}

where the shear

\begin{equation}
\sigma_{\alpha \beta} = \frac{\partial v_{\alpha}}{\partial x_{\beta}} 
+ \frac{\partial v_{\beta}}{\partial x_{\alpha}} - \frac{2}{3} 
\delta_{\alpha \beta} \nabla \cdot \bmath{v}
\end{equation}

  In these equations $\alpha$ and $\beta$ are spatial indexes while tensors are written in bold characters. For the sake of simplicity we assume $\zeta_{v} = 0$, however our code allows us also different choices. Defining

\begin{equation}
V_{i \alpha \beta} = \sum_{j=1}^{N} \frac{m_{j} \bmath{v}_{ji 
\alpha}}{\rho_{j}} \frac{\partial W_{ij}}{\partial x_{\beta}}
\end{equation}

as the SPH formulation of $\partial v_{\alpha}/\partial x_{\beta}$, the SPH equivalent of the shear is:

\begin{equation}
\sigma_{i \alpha \beta} = V_{i \alpha \beta} + V_{i \beta \alpha} - \frac{2}{3} \delta_{\alpha \beta} V_{i \gamma \gamma}
\end{equation}

A full justification of this SPH formalism can be found in \citet{b10,b11}.

In this scheme the continuity equation takes the form:

\begin{equation}
\frac{d\rho_{i}}{dt} = \sum_{j=1}^{N} m_{j} \bmath{v}_{ij} \cdot 
\nabla_{i} W_{ij}
\end{equation}

or, as we adopt, it can be written as:

\begin{equation}
\rho_{i} = \sum_{j=1}^{N} m_{j} W_{ij}
\end{equation}

which identifies the natural space interpolation of particle densities according to equation (9).

  The pressure term also includes the artificial viscosity contribution given by  \citet{b35,b36} and \citet{b42}, with an appropriate thermal diffusion term which reduces shock fluctuations. It is given by:

\begin{equation}
\eta_{ij} = \alpha_{SPH} \mu_{ij} + \beta_{SPH} \mu_{ij}^{2},
\end{equation}

where

\begin{equation}
\mu_{ij} = \left\{ \begin{array}{ll}
\frac{2 h \bmath{v}_{ij} \cdot \bmath{r}_{ij}}{(c_{si} + c_{sj}) (r_{ij}^{2} + \xi^{2})} & \textrm{if $\bmath{v}_{ij} \cdot \bmath{r}_{ij} < 0$}\\
\\
0 & \textrm{otherwise}
\end{array} \right.
\end{equation}

with $c_{si}$ being the sound speed of the ith particle, $\xi^{2} \ll h^{2}$, $\alpha_{SPH} \approx 1$ and $\beta_{SPH} \approx 2$. These $\alpha_{SPH}$ and $\beta_{SPH}$ parameters of the order of the unity are usually adopted to damp oscillations past high Mach number shock fronts developed by non-linear instabilities \citep{b6}. These $\alpha_{SPH}$ and $\beta_{SPH}$ values were also adopted by \citet{b30}. Smaller $\alpha_{SPH}$ and $\beta_{SPH}$ values, as adopted by \citet{b32}, would develop more turbulence in the disc and possibly only one shock front at the impact zone between the infalling particle stream and the returning particle stream at the disc's outer edge. In the physically inviscid SPH gas dynamics, angular momentum transport is mainly due to the artificial viscosity included in the pressure terms as:

\begin{equation}
\frac{p_{i}^{\ast}}{\rho_{i}^{2}} + \frac{p_{j}^{\ast}}{\rho_{j}^{2}} = \left( \frac{p_{i}}{\rho_{i}^{2}} + \frac{p_{j}}{\rho_{j}^{2}} \right) (1 + \eta_{ij})
\end{equation}

where $p$ is the intrinsic gas pressure.

  The advantage of an ASPH is to perform better particle interpolations ensuring a large enough number of interpolating particle neighbours. Several authors \citep{b3} have more recently adopted a criterion where the number of SPH particle neighbours for each time-step calculation is a fixed number, generally of the order of $30 \div 50$, decoupling the $h$ resolving power calculation by any physical quantity. Instead, in previous papers \citep{b36,b12,b52,b48,b31} the smoothing length $h$ has been considered a function of time by relating it to the local particle density. A spatial and temporal smoothing length together with an appropriate symmetrization concerning particle pairs have also been proposed \citep{b8,b13,b45,b46,b12,b52,b48,b31}.

  In original 3D ASPH $h_{i}$ varies in space and time. Symmetry in both $i,j$ indexes is widely adopted, where the evaluation of a symmetrized $h_{ij} = (h_{i} + h_{j})/2$ and a symmetrized Kernel $W_{ij} = (W_{ij}(h_{i}) + W_{ji}(h_{j}))/2$ are required according to:

\begin{equation}
h_{i}^{n+1} = h_{i}^{n} \left( \frac{\rho_{i}^{n}}{\rho_{i}^{n+1}} \right)^{1/3}
\end{equation}

where indexes $n$ and $n+1$ refer to time-step \citep{b13,b45,b46,b12,b52,b48,b31}. Such a choice is widely considered better than:
  
\begin{equation}
h_{i}^{n} = h_{i}^{\circ} \left( \frac{\rho_{i}^{\circ}}{\rho_{i}^{n}} \right)^{1/3}
\end{equation}

where $h_{i}^{\circ}$ and $\rho_{i}^{\circ}$ refer to initial values at time zero. Such a preference is due to the fact that because of non-linearity, instabilities can easily be produced especially in anisotropic volume changes and flow distortion \citep{b33}. Equivalently, a further equation able to compute the "new" $h$ at time-step $n+1$ from the "old" $h$ at time-step $n$ \citep{b12,b52,b48,b31} is:

\begin{equation}
h_{i}^{n+1} = h_{i}^{n} \left[ 1 + \frac{1}{3} (\nabla \cdot \bmath{v})_{i} \Delta t^{n} \right]
\end{equation}

or, by considering the continuity equation (1):

\begin{equation}
h_{i}^{n+1} = h_{i}^{n} \left[ 1 + \frac{1}{3} \left(- \frac{1}{\rho_{i}} \frac{d \rho_{i}}{dt}\right) \Delta t^{n} \right],
\end{equation}

whose integration over time gives eq. (A14). This equation is easily obtained by performing the derivative of the equation $\rho h^{3} =$ const, expressing the conservation of particle mass: \\
$d \rho/dt + 3 \rho h^{2} dh/dt = 0$, $dh/dt = - (h/3) (d \rho/dt)/ \rho$, etc.. However eqs. (22), (23) or (24) are more convenient than eq. (A14). \citet{b52} and  \citet{b48}, proposed an adaptive method splitting the 3D scheme into three 1D schemes formulating a factorized Gaussian Kernel of three 1D Gaussian components. In such a scheme a tensorial computation of SPH equations has been developed and each ASPH particle enlarges or contracts as a spheroid rather than a  spherule. They successfully applied their technique to a shock front cosmological problem where ASPH spheroids give a better shock resolution compared to typical SPH spherule without adopting any artificial viscosity term. In a further paper \citep{b48} the authors, admitting that artificial viscosity terms are necessary, especially in the momentum equation, handle such artificial viscosity terms suppressing or turning on them according to some physical circumstances (mainly in rarefaction conditions). A technique turning on/off the artificial viscosity has also been described in \citet{b43}.

  ASPH models adopt the SPH same formulation, where either:
  
\begin{equation}
\left\{ \begin{array}{ll} h_{ij} = \frac{1}{2} (h_{i} + h_{j}) \\
W_{ij} = W(r_{ij},h_{ij}),
\end{array} \right.
\end{equation}

instead of SPH $h_{i}$, and $W_{ij} = W(r_{ij},h)$ \citep{b8}, or:

\begin{equation}
\left\{ \begin{array}{ll} W_{ij} = \frac{1}{2} (W_{ij,i} + W_{ij,j}) \\
W_{ij,i} = W(r_{ij},h_{i}),
\end{array} \right.
\end{equation}

instead of SPH $W_{ij} = W(r_{ij},h)$, are adopted \citep{b13}. The second formulation is mostly more currently adopted.

  Non-isotropic ASPH \citep{b52,b48,b31} adopt an anisotropic algorithm to compute ellipsoid particle deformation and, consequently, the anisotropic smoothing length, according to the local particle concentration. Such a scheme is mainly used in simulations of 2D and 3D oblique shocks and of contact fluid surfaces. The algorithm computes the element $h_{\alpha \beta i}$, where $\alpha, \beta = x, y, z$, of the $3 \times 3$ symmetric matrix:

\begin{equation}
h_{\alpha \beta 1}^{n+1} = h_{\alpha \beta 1}^{n} \left[ 1 + \frac{0.5}{3} \left( \frac{\partial v_{i \alpha}}{\partial x_{\beta}} + \frac{\partial v_{i \beta}}{\partial x_{\alpha}} \right) \Delta t^{n} \right],
\end{equation}

where $h_{\alpha \beta i} = h_{\beta \alpha i}$, is the projection of the ellipsoid characteristic semiaxes on the cartesian axes. The eigenvectors of the matrix are the directions along the three axes of the ellipsoid and the corresponding eigenvalues are the dimensions of the ellipsoid along each axis. The determinant of the same matrix determines the normalization volume of each particle.

  The SPH conversion of eq. (A20), similarly to the SPH expression of the $\nabla \cdot \bmath{v}$ \citep{b35,b36} is:

\begin{eqnarray}
h_{\alpha \beta i}^{n+1} & = & h_{\alpha \beta i}^{n} \Bigg[ 1 + \frac{0.5 \Delta t^{n} }{3} \sum_{j=1}^{N} \frac{m_{j}}{\rho_{i}} \Big( v_{\alpha ij} \nabla_{i \beta} W_{ij} + \nonumber \\ & & v_{\beta ij} \nabla_{i \alpha} W_{ij} \Big) \Bigg].
\end{eqnarray}

\subsection{Conservative ASPH formulation}

   \citet{b45,b46} showed that energy conservation improves if $\partial/\partial h$ are introduced into both SPH momentum and energy equations. The inclusion of such terms modify substantially those equations in a non practical form. The formal difficulties were overcome by \citet{b54} who derived an effective ASPH conversion of the pressure gradient contribution in the momentum equation (eq. 2), conserving energy and entropy, according to the conservative ASPH equation:

\begin{eqnarray}
\frac{d \bmath{v}_{i}}{dt} & = & - \sum_{j=1}^{N} m_{j} \Bigg( f_{i} \frac{p_{i}^{\ast}}{\rho_{i}^{2}} \nabla_{i} W_{ij,i} + f_{j} \frac{p_{j}^{\ast}}{\rho_{j}^{2}} \nabla_{j} W_{ij,j} +  \nonumber \\ & & \Pi_{ij} \nabla_{i} W_{ij} \Bigg) + \bmath{g}_{i},
\end{eqnarray}

where $f_{i} = \left(1 + \frac{h_{i}}{3 \rho_{i}} \frac{\partial \rho_{i}}{\partial h_{i}}\right)^{-1}$, $W_{ij,i} = W(r_{ij},h_{i})$ and $\Pi_{ij}$ refers to the artificial viscosity contribution. Smoothing length $h$ was computed requiring that a fixed mass is contained within a smoothing volume: $(4 \pi /3) h_{i}^{3} \rho_{i} = M_{i,j}$ where $M_{i,j} = m_{j} N_{i,j}$ refers to the global mass of $N_{i,j}$ neighbours related to the $ith$ particle. Each particle neighbour has a $m_{j}$ mass. No further modifications to the energy equation are required. In a further paper \citep{b39} similar conclusion, as far as both SPH and XSPH methods are concerned, were reached with the aim of achieving better energy and entropy conservation.

  The $\partial \rho_{i}/ \partial h_{i}$ term is easily connected to the $\partial W_{ij}/\partial h_{i}$ by the simple relation:

\begin{eqnarray}
\frac{\partial \rho_{i}}{\partial h_{i}} & = & \sum_{j=1}^{N} m_{j} \frac{\partial W_{ij}}{\partial h_{i}},
\end{eqnarray}

  where the derivative $\partial W_{ij}/\partial h_{i}$ strictly involves also the derivative of the $h^{-3}$ in 3D as: $\partial (W_{ij} h_{i}^{-3})/\partial h_{i}$. In this scheme, the conservative ASPH conversion of the Navier-Stokes equation (eq. A7) is:
  
\begin{eqnarray}
\frac{d \bmath{v}_{i}}{dt} & = & - \sum_{j=1}^{N} m_{j} \Bigg( f_{i} \frac{p_{i}^{\ast}}{\rho_{i}^{2}} \nabla_{i}W_{ij,i} + f_{j} \frac{p_{j}^{\ast}}{\rho_{j}^{2}} \nabla_{j}W_{ij,j} + \nonumber \\ & & \Pi_{ij} \nabla_{i} W_{ij} \Bigg) + \bmath{g}_{i} + \sum_{j=1}^{N} m_{j} \Bigg( f_{i} \frac{\eta_{vi} \bmath{\sigma}_{i}}{\rho_{i}^{2}} \cdot \nabla_{i}W_{ij,i} + \nonumber \\ & & f_{j} \frac{\eta_{vj} \bmath{\sigma}_{j}}{\rho_{j}^{2}} \cdot \nabla_{j} W_{ij,j} \Bigg)
\end{eqnarray}.

  As far as the conservative ASPH energy balance equation for the total energy $E$ is concerned,

\begin{eqnarray}
\frac{d}{dt} E_{i} & = & - \sum_{j=1}^{N} m_{j} \Bigg( f_{i} \frac{p_{i}^{\ast} \bmath{v}_{i}}{\rho_{i}^{2}} \cdot \nabla_{i} W_{ij,i} + \nonumber \\
& & f_{j} \frac{p_{j}^{\ast} \bmath{v}_{j}}{\rho_{j}^{2}} \cdot \nabla_{i} W_{ij,j} \Bigg) + \sum_{j=1}^{N} m_{j} \bmath{\Omega}_{ij} \cdot \nabla_{i} W_{ij} + \nonumber \\
& & \sum_{j=1}^{N} m_{j} \Bigg( f_{i} \eta_{vi} \frac{\bmath{\sigma}_{i} \cdot \bmath{v}_{i}}{\rho_{i}^{2}} \cdot \nabla_{i} W_{ij,i} + \nonumber \\
& & f_{j} \eta_{vj} \frac{\bmath{\sigma}_{j} \cdot \bmath{v}_{j}}{\rho_{j}^{2}} \cdot \nabla_{i} W_{ij,j} \Bigg) + \bmath{g}_{i} \cdot \bmath{v}_{i},
\end{eqnarray}

where $\bmath{\Omega}_{ij}$, includes artificial viscosity terms. In conservative ASPH approach, it is easy to update the particle smoothing resolution length $h_{i}$, fixing the number of particle neighbours. In fact, according to the SPH interpolation criterion, particle concentration $n_{i} = \sum_{j=1}^{N} W_{ij}$. We remind that Kernel $W_{ij}$ is a normalized smooth function of the ratio $r_{ij}/h_{ij}$. Therefore, if $N_{neigh}$ represents the fixed number of neighbours, $h_{i} = (N_{neigh}/n_{i})^{1/3}$.

%
%__________________________________________________________________
%
%

\section{Tests}

%-------------------------------------------------- FIG B1 START
\begin{figure*}
%\hspace{25 mm}
\includegraphics[width=16cm,height=8cm]{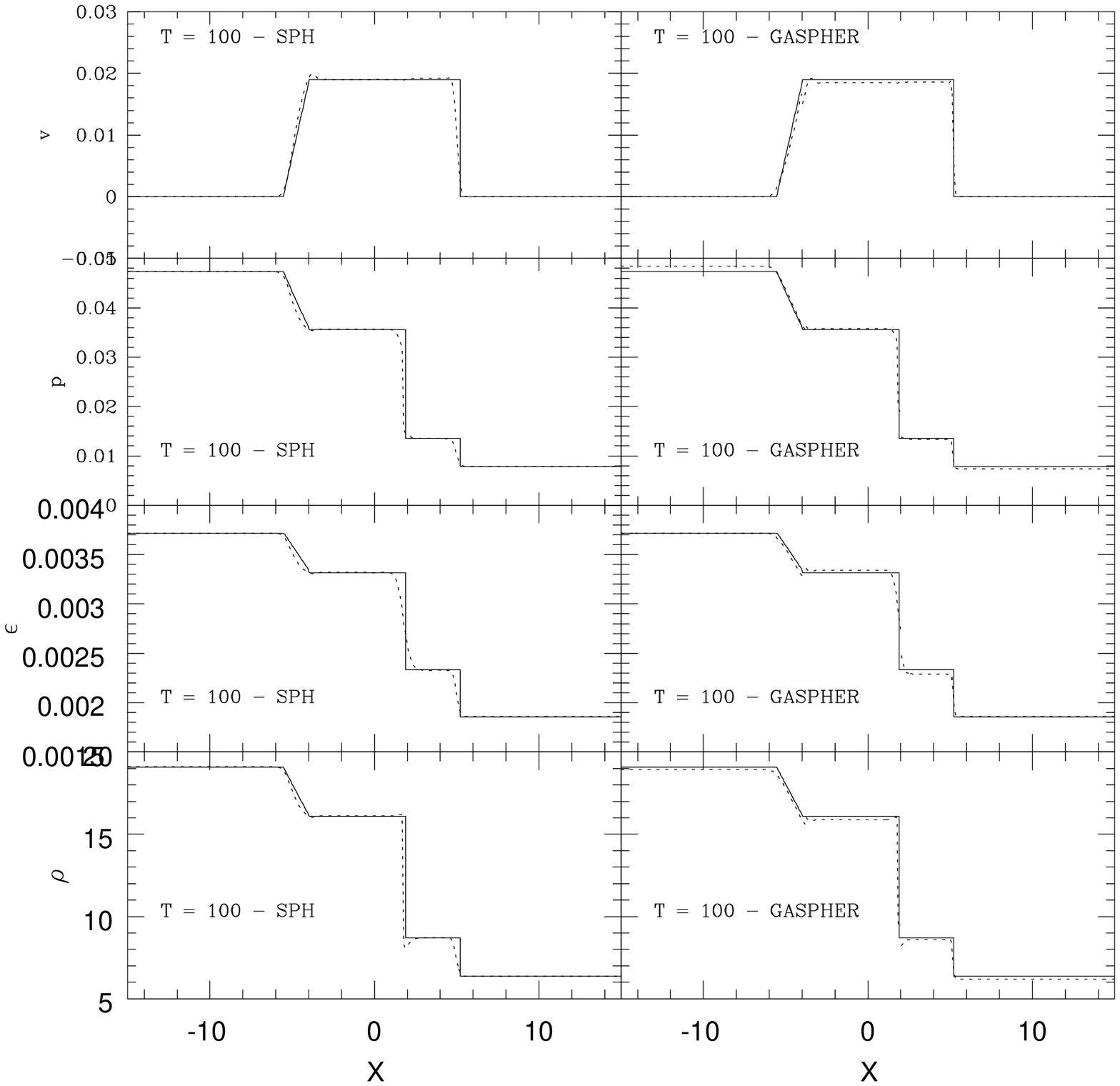}
\caption{1D shock tube tests as far as both analytical (solid line) and both SPH and GASPHER (short dashes) results are concerned. Density $\rho$, thermal energy $\epsilon$, pressure $p$ and velocity $v$ are plotted at time $T = 100$. The initial velocity is zero throughout.}
\end{figure*}
%-------------------------------------------------- FIG B1 END

%-------------------------------------------------- FIG B2 START
\begin{figure*}
%\hspace{25 mm}
\includegraphics[width=16cm,height=8cm]{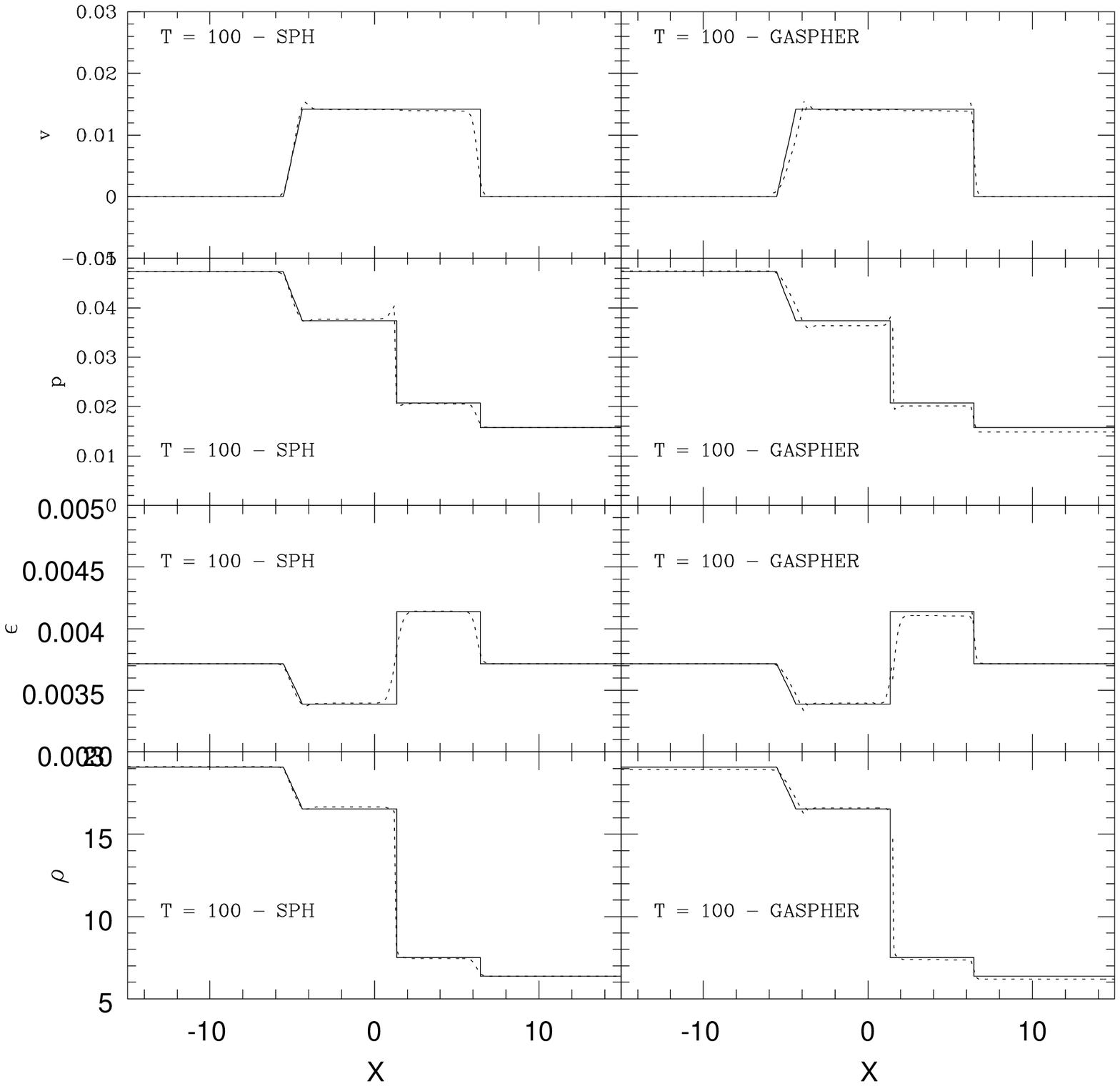}
\caption{1D shock tube tests as far as both analytical (solid line) and both SPH and GASPHER (short dashes) results are concerned. Density $\rho$, thermal energy $\epsilon$, pressure $p$ and velocity $v$ are plotted at time $T = 100$. The initial velocity is zero throughout.}
\end{figure*}
%-------------------------------------------------- FIG B2 END

%-------------------------------------------------- FIG B3 START
\begin{figure*}
%\hspace{25 mm}
\includegraphics[width=16cm,height=8cm]{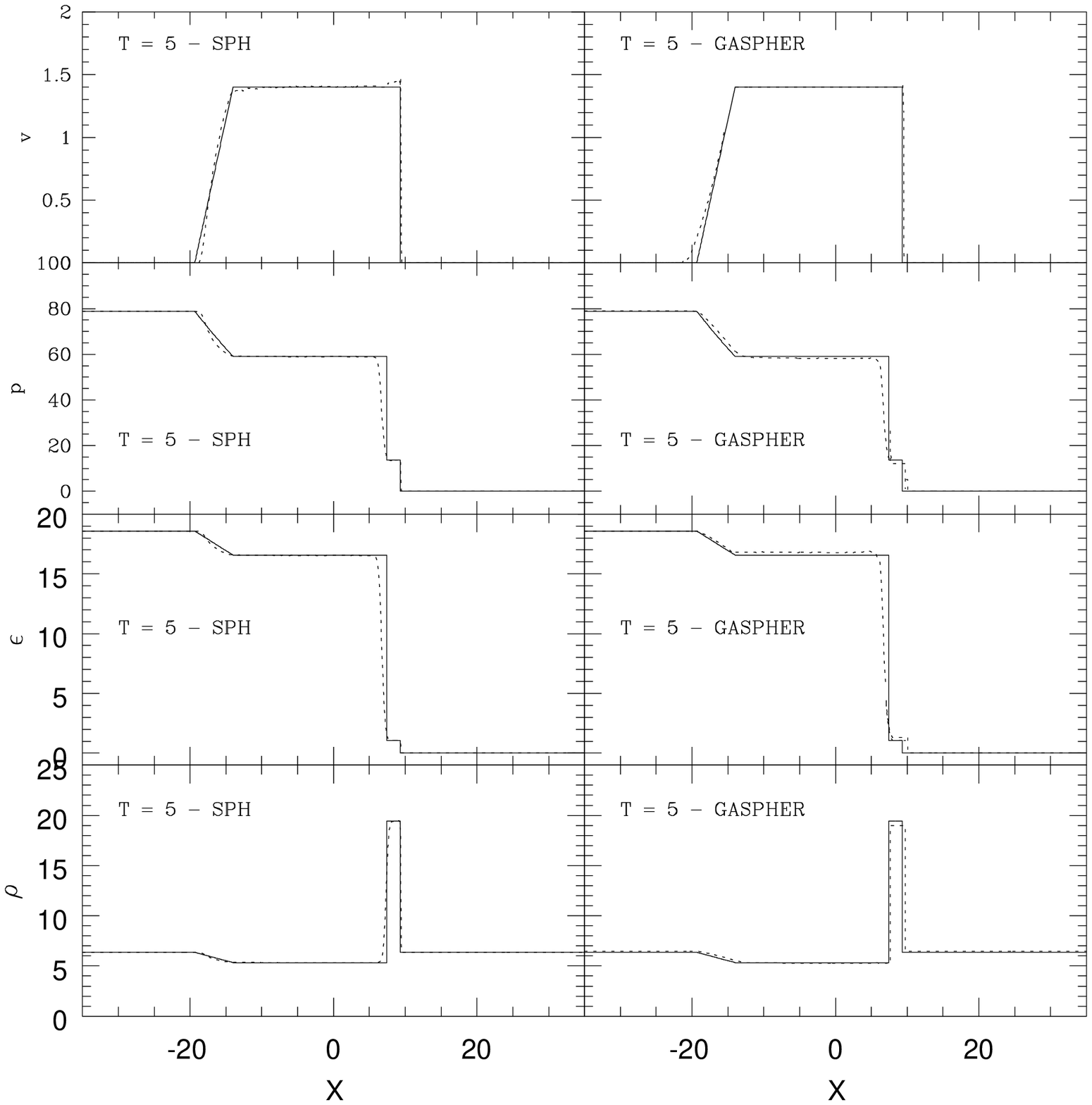}
\caption{1D blast wave tests as far as both analytical (solid line) and both SPH and  GASPHER (short dashes) results are concerned. Density $\rho$, thermal energy $\epsilon$, pressure $p$ and velocity $v$ are plotted at time $T = 5$. The initial velocity is zero throughout.}
\end{figure*}
%-------------------------------------------------- FIG B3 END

  In this section, results of some tests are here reported regarding models where either 1D shock problems, or 2D free edge, or 2D transport themes have to be taken into account to respect the argument declared in this paper. Comparison among GASPHER, SPH and ASPH numerical results are reported as well as theoretical analytical ones, whenever the theoretical analytical solution is known. The particle smoothing resolution length $h$, normally adopted throughout, is $h = 5 \cdot 10^{-2}$ (in ASPH as the initial value), but than when explicitly written. $\gamma = 5/3$ throughout. Once stated the validity of GASPHER for shock collisional modelling, a particular attention is addressed both to free edge and to radial transport results regarding the main argument of this paper.

\subsection{1D Sod shock tube tests}

  In this section a comparison of analytical and GASPHER 1D inviscid shock tube test results \citep{b63}, is made. Notice that the so called analytical solution of the 1d shock tube test is obtained through iterative procedures left-right, applying to the discontinuity the Rankine-Hugoniot "jump" solution. Figg. B1 and B2 display results concerning the particle density, thermal energy per unit mass, pressure and velocity, after a considerable time evolution at time $T = 100$. The whole computational domain is built up with $2001$ particles from $X = 0$ to $X = 100$, whose mass is different, according to the shock initial position. At time $T = 0$ all particles are motionless. $\gamma = 5/3$, while the ratios $\rho_{1}/\rho_{2} = 3$ and $\epsilon_{1}/\epsilon_{2} = 2$, and  $\rho_{1}/\rho_{2} = 3$ and $\epsilon_{1}/\epsilon_{2} = 1$ as displayed at the edges of Figg. 4 and 5, between the two sides left-right. The first $5$ and the last $5$ particles of the 1D computational domain, keep fixed positions and do not move. The choice of the final computational time is totally arbitrary, since the shock progresses in time. $v = 0$ at the beginning of each simulation. Hence, the adimensional temporal unity is chosen so that $\int_{0}^{l}dx/c_{s} = 1$. Being the sound velocity initially constant, this mathematically means $l = c_{s}$. SPH results, adopting the same initial and boundary conditions, as well as the same particle smoothing resolution length $h$, together with the analytical solutions are also displayed in the same plots.

  Our GASPHER results, are in a good comparison with the analytical solution. Discrepancies involve only $4 \div 5$ particle smoothing resolution lengths at most. This means that, GASPHER interpolations are effective in the case of shock collision case in so far as the Mach number flows regard the weak shock regimes when the Mach number ranges within $[0, 2]$ at the first instant. The decrease of the particle smoothing resolution length could improve the whole result in so far as the artificial viscosity term (depending on $h$ - eq. A12) is able to prevent particle interpenetration.

\subsection{1D Blast wave}

  Whenever in a shocktube the ratios $p_{1}/p_{1} = \epsilon_{1}/\epsilon_{2} \gg 1$, and consequently $\rho_{1}/\rho_{2} = 1$, and $v_{1} = v_{2} = 0$, such a discontinuity is called a "blast wave". Being $v = 0$ at the beginning of each simulation, the adimensional temporal unity is chosen as previously written in the 1D Sod shocktube test before. In such a situation, the Mach number spans from $0$ to $> 1$ values up to $10 \div 20$ or more at the first instant. Fig. B3 displays a comparison of SPH and GASPHER results with the so called analytical solution, after a considerable time evolution at time $T = 5$. The analytical solution is considered corrected in so far as $\rho_{1}/\rho_{2} \leq (\gamma + 1)/(\gamma - 1)$. In the blast wave test here considered, $p_{1}/p_{1} = \epsilon_{1}/\epsilon_{2} = 10^{4}$, while other spatial, initial and boundary conditions, as well as the particle spatial smoothing resolution length are identical to those chosen in the previous test. Fig. B3 displays that SPH and GASPHER results globally compare with each other and that they also compare with the analytical solution wherever $\rho_{1}/\rho_{2} \leq (\gamma + 1)/(\gamma - 1)$, that is wherever the Rankine-Hugoniot jump conditions hold. Beyond this limit, even the so called analytical solution is considered incorrect. Being $\gamma = 5/3$, the comparison is meaningful within $\rho_{1}/\rho_{2} \leq 4$. GASPHER profiles suffer of a lesser instability in those regions where particle concentration is larger, close to discontinuity profiles, where horizontal plateaus are more regular. In particular such behaviour can be addressed to the absence of any particle pair instability because of the analytical expression of the GASPHER adopted Kernel and to its radial spatial derivative.

\subsection{2D expansion of the free edge of a squared box}

  The test here discussed does not have an analytical solution. However, it is interesting because it shows how pressure forces push away the free edge of the fluid computational domain, without any explicit dissipation, according to the chosen interpolation Kernel. Being in permanent gradual expansion, any artificial viscosity contribution is statistically turned off, apart some contribution due to the shear flow close to the two marginal vertical fixed edges.

  The box is a square $4.8 \times 4.8$, having three fixed sides: two vertical unlimited sides (left - right), at $X \in [0., 0.05]$ and at $X \in [4.75, 4.80]$, and the horizontal one at the bottom from $Y = 0.$ to $Y = 0.05$, while the fourth side at $Y = 4.80$, is free to expand towards the outer space. Particles, whose mass $m_{i} = 3 \cdot 10^{-5}$ are regularly located so that their mutual separation equals $h$. The initial thermal energy is $\epsilon = 1.9188 \cdot 10^{-5}$, while the initial $v_{i} = 0$ throughout. The three fixed edges are composed of two lines of fixed particles, whose velocity is abruptly put to zero time by time. Notice that the above mentioned constraints have to be considered as geometric conditions not pertinent only to the edge particles of the square box. In this way, particles can move only toward the $Y$ direction, from $Y = 4.8$ onwards, and any particle horizontal translation is mechanically prevented.

%-------------------------------------------------- FIG B4 START
\begin{figure}
%\hspace{25 mm}
\includegraphics[width=8cm,height=8cm]{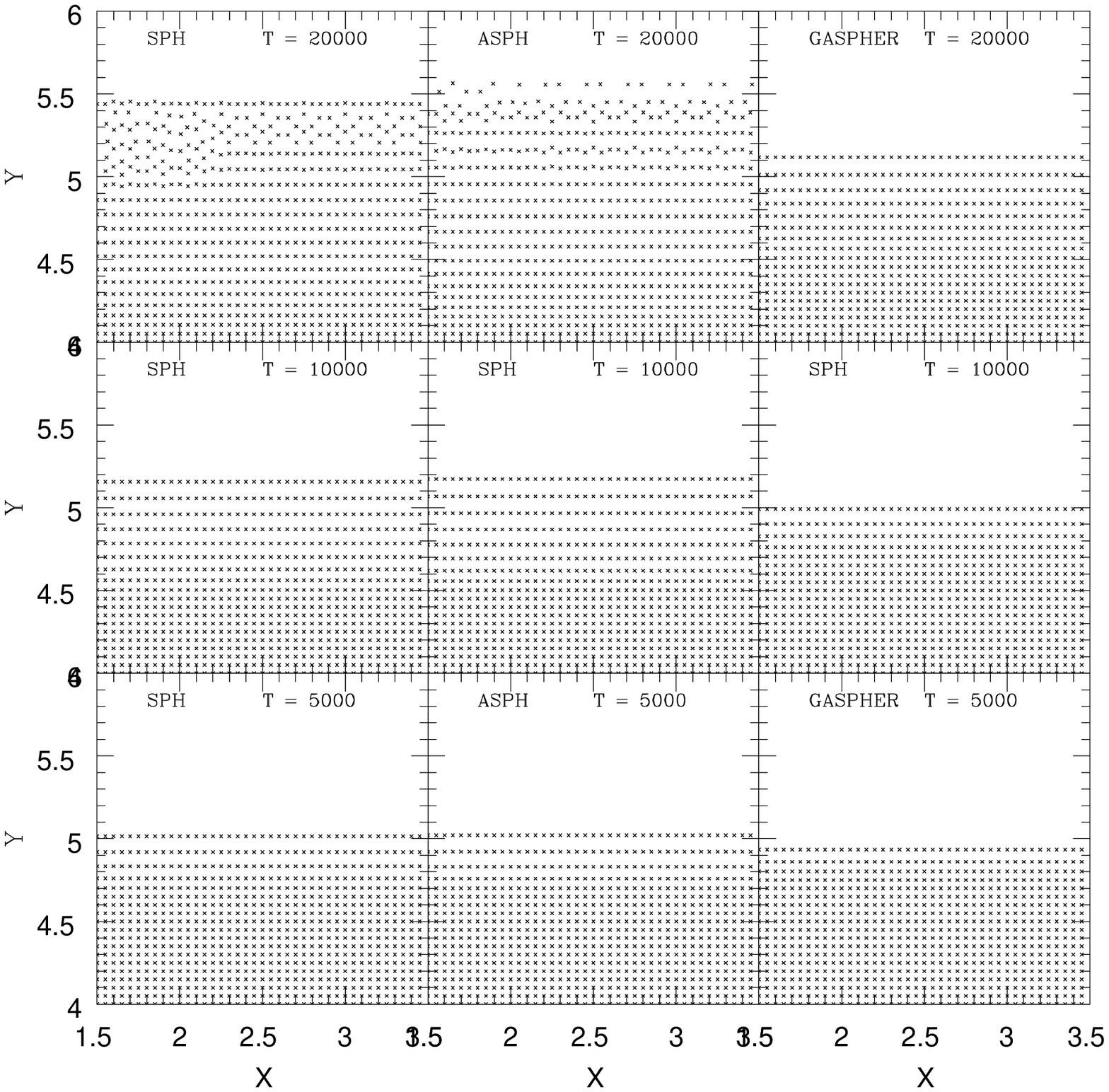}
\caption{SPH, ASPH and GASPHER $XY$ plots of the 2D expansion of a portion of the free edge on the top of a squared box. Time $T$ is shown.}
\end{figure}
%-------------------------------------------------- FIG B4 END

%-------------------------------------------------- FIG B5 START
\begin{figure}
%\hspace{25 mm}
\includegraphics[width=8cm,height=8cm]{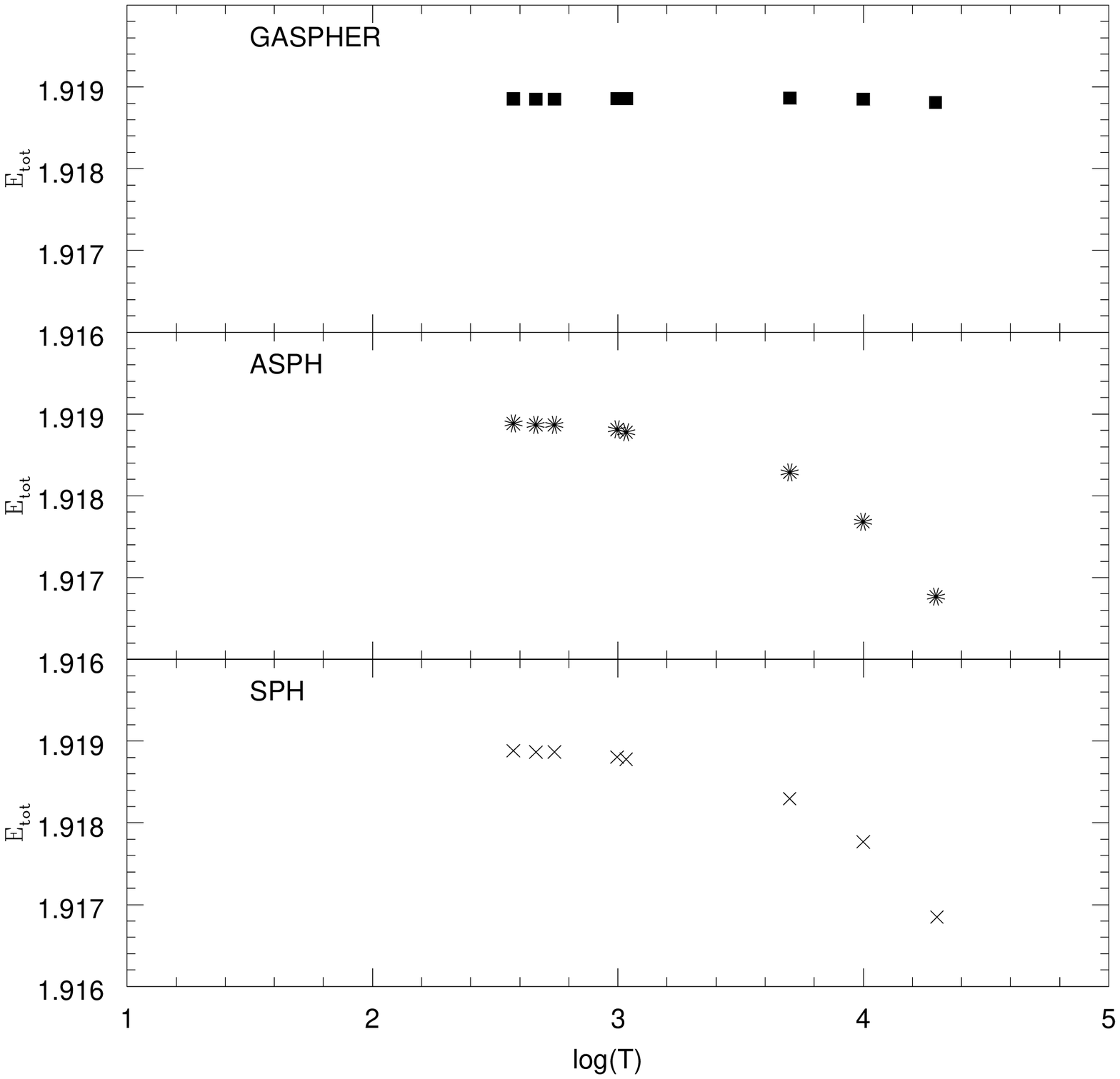}
\caption{Plots of specific total energy $E_{tot}$ averaged for each particle for the SPH, ASPH and GASPHER simulations of the 2D expansion of the free edge in a box. $E_{tot}$ values are expressed in $10^{-5}$ units. Time $T$ is reported on a logarithmic scale.}
\end{figure}
%-------------------------------------------------- FIG B5 END

  Since an analytical solution is unknown, we pay attention to the conservation of the total energy $E_{tot} = v^{2}/2 + \epsilon$ per unit mass averaged for each particle and, at the same time to the regular face of the expansion of the free horizontal edge at the top. Fig. B4 displays the advance of the free front at three selected times for the SPH, ASPH and GASPHER simulations. SPH and particularly ASPH fronts are without any doubt more advanced than the GASPHER front. This effect is the result of an incorrect computation of the pressure forces on the free edge of the computational domain as discussed in \S5. This conclusion is stressed not only by the fact that the GASPHER flow is more regular and free from defects, but also, as it is shown in Fig. B5, by the fact that the total energy per unit mass $E_{tot}$ is much better conserved than in the other two cases. As an order of magnitude, the degradation of the total energy is $\approx 10^{-8}$ for a totality of $\approx 10^{4}$ particles after a time $T \approx 2 \cdot 10^{4}$ for both SPh and ASPH. This involves that on a single particle, $dE/dt \approx 5 \cdot 10^{-13}$. Instead, in GASPHER this energy degradation is $\sim 10^{2}$ times smaller. This implies that the choice of the interpolation Kernel is crucial in the conservation of prime integrals.

\subsection{2D radial spread and migration of a Keplerian annulus ring}

  The 2D radial spread and migration of an isothermal Keplerian annulus ring is widely described in \citet{b85} in the case of a constant physical viscosity $\nu$. At time $T = 0$, the surface density, as a function of the radial distance $r$, is described by a Dirac $\delta$ function: $\Sigma (r,0) = M \delta (r - r_{\circ})/2 \pi r_{\circ}$, where $M$ is the mass of the entire ring and $r_{\circ}$ is its initial radius. As a function also of time, the surface density is computed via standard methods as a function of the modified Bessel function $I_{1/4} (z)$:

%-------------------------------------------------- FIG B6 START
\begin{figure}
\resizebox{\hsize}{!}{\includegraphics[clip=true]{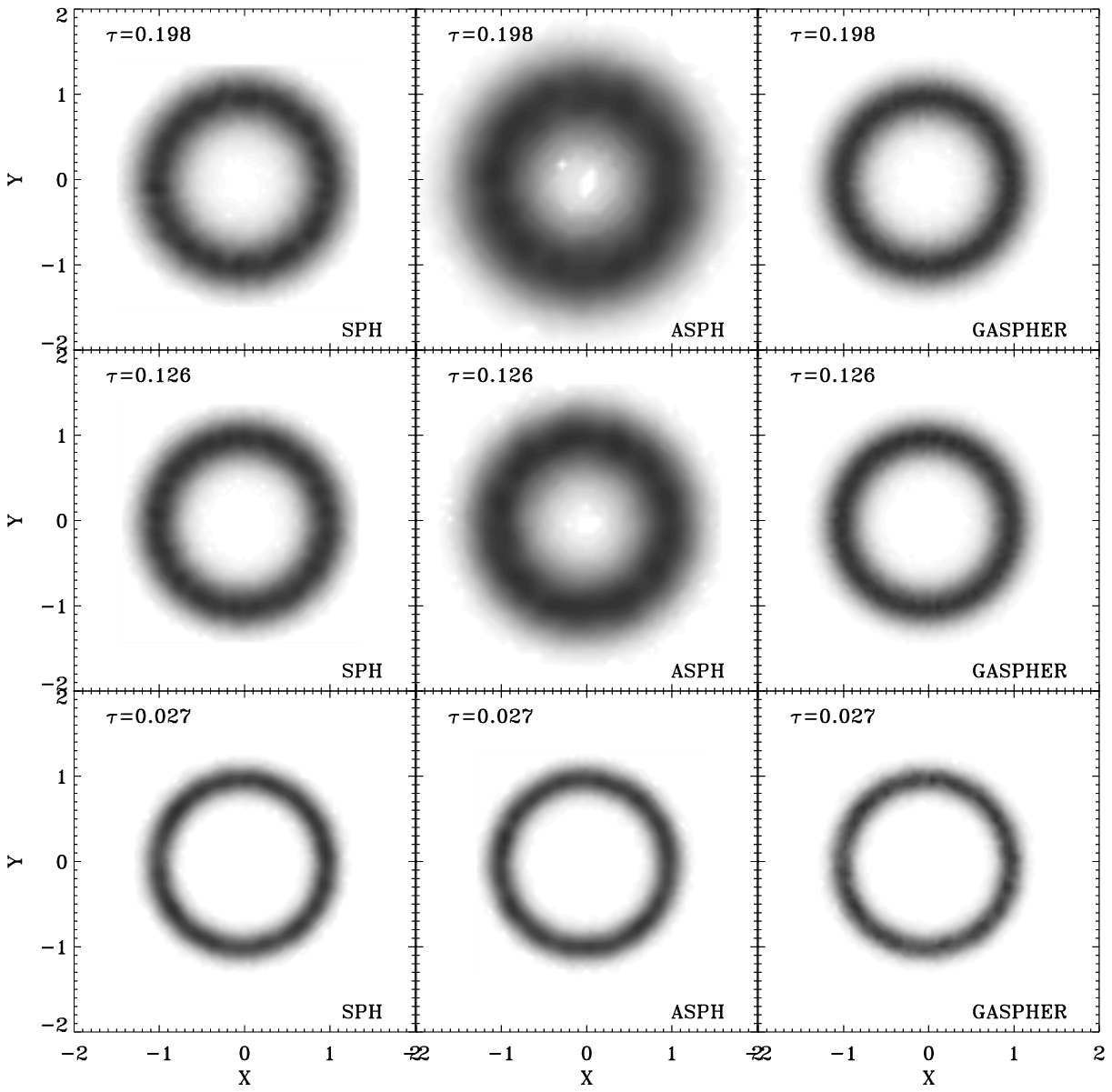}}
\caption{$XY$ plots of ring density contour maps. Times are reported for each configuration (SPH or ASPH or GASPHER).}
\end{figure}
%-------------------------------------------------- FIG B6 END

%-------------------------------------------------- FIG B7 START
\begin{figure}
\resizebox{\hsize}{!}{\includegraphics[clip=true]{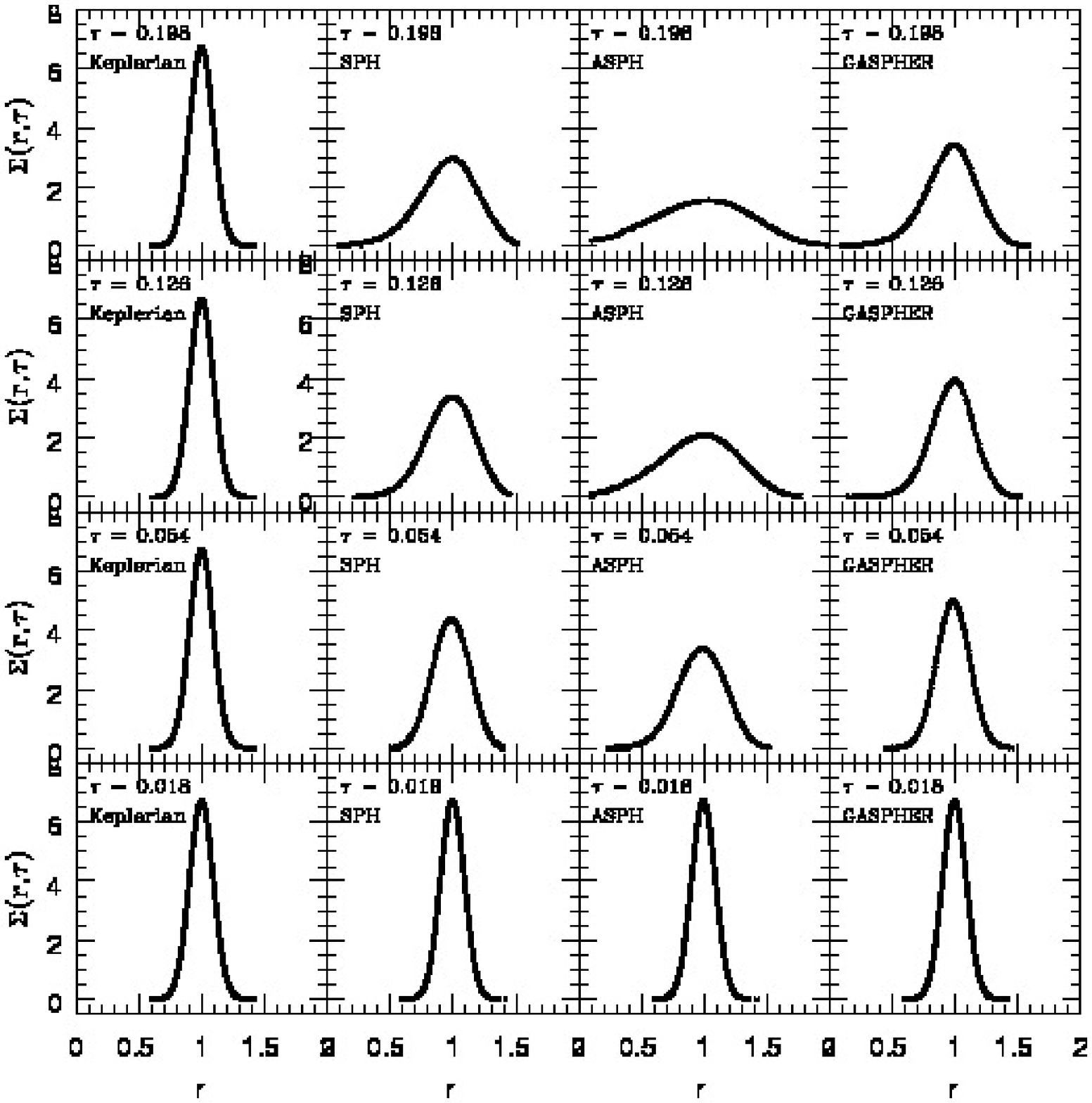}}
\caption{Surface density $\Sigma (r, \tau)$ in $10^{-11}$ units as a function of radial distance from initial configurations at $\tau = 0.018$. Subsequent snapshots are reported for each configuration both SPH or ASPH or GASPHER.}
\end{figure}
%-------------------------------------------------- FIG B7 END

\begin{equation}
\Sigma (x, \tau) = \frac{M}{\pi r_{\circ}^{2}} \tau^{-1} x^{-1/4} e^{- \frac{1 + x^{2}}{\tau}} I_{1/4} (2x/\tau),
\end{equation}

where $x = r/r_{\circ}$, $\tau = 12 \nu T/r_{\circ}^{-2}$. $\int_{S} \Sigma (x, \tau) dS = 2 \pi \int \Sigma (x, \tau) dr =$ const equals the annulus mass throughout. Time is normalized so that $T = 1$ is the Keplerian period corresponding to the ring at $r_{\circ} = 1$. Examples of SPH viscous spread on this argument can be found in \citet{b11,b89,b87}, as well as in \citet{b88} in SPH physically inviscid hydrodynamics on the basis that the shear dissipation in non viscous flows can be compared to physical dissipation \citep{b34,b44,b47}. In particular an exhaustive comparison can also be found in \citet{b20,b56}.

  In a non viscous particle Lagrangian fluid dynamics, any deviation from the initial strictly Keplerian kinematics is incorrectly due to the activation of artificial viscosity dissipation in the shear flow when two particles approach each other. This is an unavoidable consequence of the fact that dissipation is currently used to handle the direct head-on collision between pair of particles. To establish whether the adopted Kernel has a significant role in the spatial transport phenomena, any gas pressure force component must be removed leaving active only the artificial viscosity dissipation in the momentum equation in a strictly isothermal fluid dynamics. Thus, the whole flow should keep its Keplerian behaviour because pressure forces are artificiously erased, in so far as artificial viscosity dissipation stays inactive. In this test, the two marginal edges of the annulus (the inner and the outer ones) are considered as FE boundaries. Adopting the same artificial viscosity formulation and the same parameters ($\alpha = 1$ and $\beta = 2$ - see App. A) both for SPH and for ASPH and for GASPHER simulations, our aim is to check which technique shows the smaller deviation from the initial Keplerian tight particle distribution in isothermal conditions, keeping constant both the sound velocity and the specific thermal energy. SPH-derived techniques turns on the artificial viscosity dissipation whenever two close particles approach with each other. This happens also for shear flows. However for inviscid ideal shear flows this is an incorrect result without any gas compression.

  A significant comparison of GASPHER to SPH and to ASPH is displayed in Fig. B6, where $XY$ density contour map plots are shown at the same $\tau$. The radial distributions of surface density are displayed in Fig. B7, according to the restricted hypotheses of the standard mechanism of physical dissipation (constant dissipation, zero initial thickness). As in \citet{b89,b87}, the initial ring radius is at $r_{\circ} = 1$, whose thickness is $\Delta r = 0.5$, is composed of $40000$ equal mass ($m_{i} = 2.5 \cdot 10^{-15}$) pressureless Keplerian ($\bmath{v} = \bmath{v}_{Kepl}$, $\nabla \cdot \bmath{v} = 0$ at $T = 0$) SPH particles, with $h = 9 \cdot 10^{-2}$, with $c_{s} = 5 \cdot 10^{-2}$, and with initial density radial distribution corresponding to the analytical solution at time $T$, whose $\tau = 0.018$. To this purpose, a random number generator has been used. The central accretor has mass normalized to $M = 1$. The kinematic shear dissipation is estimated \citep{b34,b44,b47,b20,b56} as $\nu \approx c_{s} h$.

  GASPHER radial spread is without any doubt the narrower one, while the ASPH one is naturally the larger because of the increasing particle smoothing resolution length $h$ affecting the artificial viscosity analytical expression. For this reason, its physical dissipation counterpart $\nu \approx c_{s} h$ cannot be kept constant even preventing any disc heating. Only in the case of ASPH modelling, the initial $h$ value is assumed to compute $\tau$. Notice that these results are obtained according to the correlation $\nu \approx c_{s} h$ \citep{b34} in the expression where $\tau = 12 \nu T/r_{\circ}^{-2}$, which appears as the most appropriate. In fact, considering $\nu \approx 0.1 \alpha_{SPH} c_{s} h$ \citep{b44,b47} with $\alpha_{SPH} \approx 1$, it is necessary a time $T$ ten times longer to get the same $\tau$. This involves a larger annulus spread as far as the numerical results are concerned. According to this results, the Kernel choice is determinant also in the generation of kinematic deviations from the initial Keplerian distribution due to the incorrect SPH dissipation because of the particle shear approaching in the non viscous ideal flows.

  Notice that the density radial distribution, as far as the ASPH modelling is concerned, better fits the spread of the theoretical radial distribution (here not represented). This is a fair result in so far as we are interested in determining which artificial dissipation, coupled with the choice of the interpolation Kernel, determines a density radial profile, to be compared with the theoretical one, when a physical dissipation $\nu$ is considered. However, this is another aspect, regarding the study of either the physical dissipation in a viscous fluid dynamics or its artificial numerical dissipation counterpart in a non viscous approach, which is far from the scope of the test here proposed.

  Circular rings, appearing in Fig. B6 for both numerical schemes, are an unavoidable effect due to the Lagrangian particle-based technique, as discussed in \citet{b89,b87}.

%\end{appendix}

\label{lastpage}

\end{document}